\documentclass[11pt]{article}
\usepackage{emulateapj}
\usepackage{flushrt,natbib}
\usepackage{epsfig,journals}
\newcommand{\sigmath}{\sigma_{\rm th}}
\newcommand{\sigmaNT}{\sigma_{\rm NT}}
\newcommand{\Sigmacl}{\Sigma_{\rm cl}}
\newcommand{\Msun}{M_\odot}
\newcommand{\pc}{{\rm pc}}
\newcommand{\AU}{{\rm AU}}

\newcommand{\yr}{{\rm yr}}
\newcommand{\Mbe}{M_{\rm BE}}
\newcommand{\Rbe}{R_{\rm BE}}
\newcommand{\ceff}{c_{\rm eff}}

\newcommand{\fF}{f_F}

\newcommand{\Sigsph}{\Sigma_{\rm sph}}
\received{2004 August 26}\accepted{2005 May 6}
\begin{document}

\title{Protostellar Disks: Formation, Fragmentation, and the Brown
  Dwarf Desert} 

\author{Christopher D. Matzner$^{a}$ and Yuri Levin$^b$}
\affil{$^a$Department of Astronomy \& Astrophysics\\
$^b$Canadian Institute for Theoretical Astrophysics \\ University of
Toronto, 60 St.~George Street, Toronto, ON M5S 3H8, Canada}
\begin{abstract}
We argue that gravitational instability of typical protostellar disks
is not a viable mechanism for the fragmentation into multiple systems
-- binary stars, brown dwarf companions, or gas giant planets --
except at periods above roughly 20,000 years.  Our conclusion is based
on a comparison between prior numerical work on disk self-gravity by
Gammie (2001) with our own analytical models for the dynamical and
thermal state of protostellar disks.  For this purpose we first
develop a simple theory for the initial conditions of low-mass star
formation, accounting for the effect of turbulence on the
characteristic mass, accretion rate, and angular momentum of collaping
cores.  We also present formulae for the probability distribution of
these quantities for the case of homogenous Gaussian turbulence.
However, our conclusions are not sensitive to this parameterization.

Second, we examine the criterion for fragmentation to occur during
star formation, concentrating on the self-gravitational instabilities
of protostellar accretion disks in their main accretion phase.
Self-gravitational instabilities are strongly dependent on the thermal
state of the disk, and we find that the combination of viscous heating
and stellar irradiation quenches
fragmentation due to Toomre's local instability.  Simulations by
Matsumoto \& Hanawa (2003), which do not include detailed thermal
evolution, predict fragmentation in an early phase of collapse.  But,
fragments born in this phase are on tight orbits and are likely to
merge later due to disk accretion.  Global instability of the disk may
be required to process mass supply, but this is also unlikely to
produce fragments.  We conclude that numerical simulations which
predict brown dwarf formation by disk fragmentation, but which do not
account for irradiation, are unrealistic.  Our findings help to
explain the dearth of substellar companions to stellar type stars: the
brown dwarf desert.
\end{abstract}

\section{Introduction}\label{intro}
At birth, stars accumulate their material through disks that are well
known to be dense and massive.  Both of these properties make
protostellar accretion disks susceptible to self-gravitational
instability \citep[e.g.,][]{1964ApJ...139.1217T}.  Indeed, simulations
of protostellar disk accretion and evolution like those of
\cite{1990ApJ...358..515L} typically find that they approach or cross
the threshold for instability, either because of a high disk density
or because of a finite disk mass.  When this occurs, disk motions
caused by self-gravity become important and perhaps dominant sources
of angular momentum transport \citep{1978AcA....28...91P,
1987MNRAS.225..607L, 1991MNRAS.248..353P, 1989ApJ...347..959A,
1994pad..confR...9T, 1994ApJ...436..335L,2001ApJ...553..174G}.
Indeed, they may be required for disks to process mass accretion.

Disk self-gravity is of interest not only as a transport mechanism,
but also because it may cause the disk to fragment into 
self-gravitating bodies that can become stellar, substellar, or
possibly planetary companions to the star being formed. 

If fragments form sufficiently early and survive as most of the
stellar mass is accumulated, they stand a chance of acquiring mass
comparable to that of the primary object \citep{1994MNRAS.269..837B,
2003ApJ...595..913M}.  Alternatively, fragments that survive but fail
to accumulate material may wind up with brown dwarf or planetary
masses at the end of accretion, depending partly on where in the disk
they form.  The possibility that gas giant planets form from
fragmentation of a gaseous disk was suggested by
\cite{1978M&P....18....5C} and advocated by Boss
\citep[e.g.,][]{1997Sci...276.1836B}.  It has been investigated
numerically by \cite{2002Sci...298.1756M}, for example.  However, numerical
investigations are difficult, both because of the high resolution
required to attain convergence in self-gravitating simulations of
cooling gas \citep{1997ApJ...489L.179T}, and because of the importance
of thermal evolution to disk fragmentation
\citep{2000ApJ...529..357N,2001ApJ...553..174G,2003ApJ...590.1060P,
2004astro.ph..6469R}.  We return to disks' thermal evolution below.

The possibility that disk instability can produce stellar or substellar
companions makes it relevant to the problem of binary and multiple
star formation, and to the origin of the stellar initial mass
function.  
The binary problem is reasonably well-posed, since the prestellar
cores have been mapped \citep[e.g.,][]{1989ApJS...71...89B} and binary
stars are also well studied.  Indeed, binaries' angular momenta are
similar to cores' \citep{1991A&A...248..485D,2000ApJ...528..325B}, a
fact that explains some trends in binary properties
\citep{2004ApJ...600..769F}.  However the mechanism by which a core
fragments is currently unclear.  It is not understood, for instance,
why some stars -- albeit less than half -- form on their own. 

\cite{2002ARA&A..40..349T} has reviewed theories for the origin of
binary stars, and identified three leading mechanisms: ``prompt''
fragmentation in a sheet formed from the quasi-homologous collapse of
a cold, rotating cloud; fission of a rapidly rotating and contracting
protostar, and self-gravitational instabilities of accretion disks.
(\citeauthor{1953ApJ...118..513H}'s 1953 proposal of opacity limited
fragmentation during collapse appears unlikely, due to pressure
gradients caused by central concentration of the initial state.)  One
should add to Tohline's list {\em turbulent} fragmentation, in which
nonlinear perturbations in the initial state engender fragments
directly; see \cite{2002ApJ...576..870P}, \cite{2003RMxAC..15...92K},
\cite{2001ApJ...556..837K}, and \cite{2003MNRAS.339..577B}.

This paper will consider axisymmetric models of low-mass star
formation.
Given the turbulent character of the molecular clouds from which stars
form, one might wonder how relevant axisymmetric models can possibly
be to stellar fragmentation.  However, the energy in this turbulence
is concentrated in its longest wavelengths
\citep{1981MNRAS.194..809L}.  The direct progenitors of single and
binary stars are the dense molecular cores which are supported
significantly by thermal pressure, unlike the larger structures that
contain them -- their parent cloud or the cloud substructures known as
clumps -- which are supported entirely by turbulence and magnetic
fields \citep{PPIII}.  On the scale of an individual prestellar core,
perturbations with longer wavelength appear as translational motion,
shear, and overall compression.  All of these can be treated with an
axisymmetric model in the center of mass frame.  Motions with smaller
wavelengths constitute turbulence within the core and contribute to
turbulent fragmentation but are weaker in general and are suppressed
by ion-neutral friction \citep{1983ApJ...270..511Z}. For these
reasons, and for analytical convenience, we will concentrate on
axisymmetric initial conditions.

In this paper we develop a simple prescription for the initial
conditions of star formation and examine the susceptibility of the
predicted protostellar accretion disks to several modes of
gravitational fragmentation.  We begin by modeling the effect of
molecular cloud turbulence on the masses, radii, and rotation rates of
the prestellar molecular cores (\S~\ref{Inits}).  We then address the
``prompt'' instability that occurs early in the collapse
(\S~\ref{Impulsive}), using criteria developed by the numerical survey
of \cite{2003ApJ...595..913M}.  This analysis suggests that
fragmentation occurs quite frequently in the early collapse. However,
we argue that
this may be an artifact of approximations made in the thermal
evolution of core material, and that the fragments so formed are
destined to
merge during subsequent accretion. 

We then turn (\S~\ref{Instab}) to fragmentation due to
\cite{1964ApJ...139.1217T}'s instability. Since this is governed by
the local density, we will call it ``local fragmentation''.  Here we
must discriminate between conditions in which Toomre's instability
saturates, inducing angular momentum transport
\citep{1978AcA....28...91P, 1987MNRAS.225..607L,1991MNRAS.248..353P,
1994ApJ...436..335L, 1996ApJ...456..279L}, and conditions that lead to
local fragmentation.  For this we rely on \cite{2001ApJ...553..174G},
who identifies the boundary between saturated ``gravitoturbulence''
and fragmentation in his simulations of razor-thin accretion disks.
Although protostellar disks are hardly razor-thin, Gammie's criterion
is a useful litmus test for fragmentation.  It demonstrates that the
outcome of gravitational instability is sensitive to the thermal
evolution of the disk midplane, which we model first by treating only
viscous heating alone in \S \ref{self-luminous}.  This analysis
indicates the possibility of fragments forming with periods greater
than about 
460 years.  However when we account for irradiation from the central
star in \S \ref{irrad}, we will demonstrate that this quenches local
fragmentation in the main accretion phase, at least for periods up to
$\sim 1.2\times10^4$ years.

Lastly we treat instability that arises from the finite disk mass
\citep{1989ApJ...347..959A,1994pad..confR...9T}.  This sets in if
other transport mechanisms fail to flush the disk, and
\cite{1994MNRAS.269..837B} suggests it it may lead to binary
fragmentation.  However, current evidence from numerical simulations
\citep{1998ApJ...504..945L,1996ApJ...456..279L} shows that the
instability would always redistribute the disk material to quench
itself, and the fragmentation would never occur via this channel.

We draw conclusions in \S \ref{conclusions} about the
implications of this work for understanding the brown dwarf desert,
i.e., the observed underrepresentation of subsolar companions to solar
type stars.

\section{Star Formation: Initial Conditions} \label{Inits}

Prior to their collapse, molecular cores are partially supported by
thermal pressure and are confined by the hydrostatic pressure of their
parent cloud and permeated by its turbulent motions.  Combined with
the scalings known to apply to molecular cloud turbulence, these
statements suffice to predict, if only crudely, the initial conditions
for star formation. See also \cite{Crete99} and
\cite{2003ApJ...585..850M}.

First, consider the structure of cores prior to their collapse. In the
absence of non-thermal pressure from magnetic fields, turbulence, or
rotation, these are Bonnor-Ebert \citep[BE;][]{1956MNRAS.116..351B} spheres, 
confined by some external pressure $P$.  They
exhibit a sequence of increasing self-gravity and central
concentration, leading up to a critical state whose mass and radius satisfy
\begin{equation}\label{BEspheres}
\Mbe = 1.18 \frac{\sigmath^4}{G^{3/2} P^{1/2} } 
~~~~{\rm and } ~~~~
\frac{\Mbe}{\Rbe} = 2.43 \frac{\sigmath^2}{G}, 
\end{equation}
where $\sigmath$ isothermal sound speed.  Molecular line cooling and
dust radiation typically conspire to maintain $\sigmath \simeq 0.2$
km/s, corresponding to temperatures $T\simeq 10$ K for molecular gas
\citep{1993ApJ...418..273F}; however one can imagine other
possibilities, such as star formation in strongly heated or
metal-free gas, in which $T$ could be significantly higher.

Because non-thermal forms of pressure also contribute, it is
reasonable to replace $\sigmath$ with an effective sound speed $\ceff$
derived below. The Bonnor-Ebert parameters $\Mbe$ and $\Rbe$ provide
fiducial values for the core mass and radius, although the actual
values ($M_c$ and $R_c$, say) differ because $\ceff>\sigmath$.  We
show below that thermal and nonthermal contributions are comparable in
these objects.

\cite{1993ApJ...406..528G} and \cite{2000ApJ...543..822B} have shown
that the observed rotation rates of molecular cores are consistent
with the turbulent velocity fields of their parent clouds, if
extrapolated to the sizes of individual cores.  It follows that
$\sigmaNT$ for an individual core can be predicted the same way.

Unlike cores, molecular clouds are supported by a triumvirate of
turbulent velocity, turbulent magnetic fields (which together compose
$\sigmaNT$) and magnetic field pressure, with $\sigmath$ being
entirely negligible on scales above $R_c$.  Clouds are famously
reported to obey a line width-size relation, noted first by
\cite{1981MNRAS.194..809L} and refined by \cite{1987ApJ...319..730S},
in which
\begin{equation} \label{Larson} 
\sigmaNT(R)^2 \simeq 0.70 G \Sigmacl R,  
\end{equation} 
corresponding to a state of strong self-gravity \citep{Crete99}. 
In eq.~(\ref{Larson}), $R$ refers to the scale on which $\sigmaNT$ is
measured -- on the core scale, $\sigmaNT$ would be $\sigmaNT(R_c)$.

The column density of the parent cloud or molecular clump is
$\Sigmacl$; \cite{1987ApJ...319..730S} find, using a virial analysis,
that for giant molecular clouds $\Sigmacl$ is remarkably constant
around the value 165 $\Msun/\pc^2$.  However, $\Sigmacl$ does depend
on context: in M33, for example, \cite{2003astro.ph..7322R} derive
$\Sigmacl\simeq 120 \Msun/\pc^2$; in the LMC,
\cite{1998ApJ...498..735P} derive $\Sigmacl \sim 700 \Msun/\pc^2 ~(\pm
0.3 {\rm dex})$; and in Galactic clumps forming massive stars
\citep{1997ApJ...476..730P,2003ApJ...585..850M} or in extragalactic
starburst regions \citep[e.g.,][]{1991ApJ...366L...1S,1991ApJ...366L...5S},
$\Sigmacl\sim 3000-5000 \Msun/\pc^2$.  

The confining pressure for a core must derive from the mean hydrostatic
pressure of the parent clump or cloud; this is \citep{Crete99}
\begin{equation}\label{eq:P} 
P = 0.51 f_p G \Sigmacl^2.  
\end{equation} 
where we have included $f_p$ to account for variations in clump or
cloud structure (0.51 refers to a spherical cloud with $\rho\propto
1/r$ and no embedded stars) and for fluctuations around the mean value
due to turbulence or location within the cloud.  With equations
(\ref{BEspheres}), this implies
\begin{eqnarray}\label{MbeRbeEvald}
M_{\rm BE} = 1.65 \frac{\sigmath^4}{f_p^{1/2} G^2 \Sigmacl } =
\frac{0.70}{f_p^{1/2}} \frac{165\Msun\pc^{-2}}{\Sigmacl} T_{\rm
eff,10}^2 \Msun, \nonumber\\ R_{\rm BE} = 0.68 \frac{\sigmath^2
}{f_p^{1/2} G \Sigmacl} = \frac{0.034}{f_p^{1/2}}
\frac{165\Msun\pc^{-2}}{\Sigmacl} T_{\rm eff,10} \pc
\end{eqnarray} 
where $T_{\rm eff,10}$ is the temperature for which $\sigmath =
\ceff$, measured in units of 10 K.  

We wish to highlight two points about Bonnor-Ebert spheres confined by
the ambient pressure (equations [\ref{BEspheres}] and [\ref{eq:P}])
and bathed in turbulence described by equation (\ref{Larson}).  First,
note that the column density of the core is similar to that of the
parent cloud or clump:
\begin{equation}\label{SigmaBE}
{\Mbe\over\pi \Rbe^2} = 1.1 f_p^{1/2} \Sigmacl, 
\end{equation}
independently of $\sigmath$.  This results from the fact that both the
core and the region in which it is embedded are self-gravitating
objects whose pressures follow equation (\ref{eq:P}), and also from
the fact that the core's mean pressure is proportional to its surface
pressure.  \cite{2003ApJ...585..850M} find a relation quite similar to
eq. (\ref{SigmaBE}) for turbulence-supported cores.

Second, compare $\sigmaNT(\Rbe)$ to $\sigmath$:
\begin{equation}\label{MachBE} 
\frac{\sigmaNT(\Rbe)}{\sigmath} = \frac{0.69}{f_p^{1/4}}.
\end{equation} 
independently of $\Sigmacl$.  Equation (\ref{Larson}) implies that any
cloud substructure whose column density is comparable to $\Sigmacl$
(i.e, any self-gravitating substructure) is suffused with turbulence
at the virial level.  Extrapolated to the scale of a Bonnor-Ebert
sphere, which is assumed to be supported by thermal pressure, this
implies $\sigmaNT\simeq\sigmath$ as in equation (\ref{MachBE}).
Heuristically, the hydrostatic cloud or clump pressure $P$, which is
nonthermal on scales above $\Rbe$, must be continuous with the 
core pressure, which is largely thermal.

Equation (\ref{MachBE}), in turn, implies that nonthermal motions must
be included in the support of cores.
To be specific, we adopt\footnote{
 The coefficient of $\sigmaNT^2$ in eq. (\ref{ceff}) depends on how 
 $\ceff$ is used.  For Alfv\'en waves treated as a gas with adiabatic
 index 3/2 and polytropic index 1/2, this coefficient is 3/2 for $P =
 \rho \ceff^2$ \citep{1995ApJ...440..686M}, 2/3 for $\Mbe =
 1.18\ceff^3/[G^{3/2} \rho({\rm edge})^{1/2}]$ \citep{MH99}, and 1.06
 for $\Mbe = 1.18 \ceff^4/(G^{3/2} P^{1/2})$. 
}
\begin{equation}\label{ceff}
\ceff^2 = \sigmath^2 + 1.06\sigmaNT^2
\end{equation}
motivated
by \cite{MH99}'s analysis of the mass of regions supported
by Alfv\'en waves \citep[treated in the WKB
approximation:][]{1995ApJ...440..686M}.  With this,  equation
(\ref{MachBE}) is solved if 
\begin{equation} \label{MachCore}
\sigmaNT \simeq {0.98\over f_p^{1/2}} \sigmath, ~~~~~
T_{\rm eff} \simeq {2.02\over f_p^{1/2}} T_c,~~~~~~
({\rm cores})
\end{equation}
which are power-law fits to the algebraic solution (good to $20\%$ for
$0.5<f_p<3$).  Cores are therefore somewhat larger than objects
lacking turbulent support:
\begin{eqnarray}\label{coresWithTurb}
M_c \simeq \frac{4.1}{f_p} \Mbe  = \frac{2.8}{f_p^{3/2}} \frac{165\Msun\pc^{-2}}{\Sigmacl} T_{10}^2 ~\Msun, \nonumber\\
 R_c \simeq \frac{2.0}{f_p^{1/2}} \Rbe = \frac{0.069}{f_p}  
\frac{165\Msun\pc^{-2}}{\Sigmacl} T_{10} ~\pc 
\end{eqnarray} 
These values are more consistent with observations than $\Rbe$ and
$\Mbe$, which points to the presence of turbulent support.  The fact
that $M_c$ exceeds the mean stellar mass is no concern:
\cite{2000ApJ...545..364M} have shown that low-mass stars form at an
efficiency $\varepsilon \simeq 25\%-80\%$, due to the action of
protostellar winds.  Bear in mind, moreover, that there is a mass
spectrum of cores going to larger and smaller masses, which generally
resembles in its shape the stellar initial mass function
\citep{1998A&A...336..150M}. 

It is also worth stressing that the objects we call ``cores'' are only
the smallest and most thermally-supported self-gravitating regions
identifiable within molecular clouds.  The clumps that form clusters
of stars, and clouds themselves, have $\sigmaNT\gg \sigmath$ and
masses much greater than $M_c$.

We have not yet specified the distribution of the variable $f_p$ that
determines cores' bounding pressures in comparison to the cloud or
clump mean pressure.  This is a product of several factors: the
distribution of pressure within a cloud; the distribution of pressure
with time due to turbulent fluctuations; and the likelihood of forming
an unstable core, which presumably biases the distribution to higher
pressures.  On the other hand, the $f_p$ distribution is strongly
constrained if one posits that stars always form from marginally
stable, thermally-supported cores whose masses are given by equation
(\ref{coresWithTurb}).  In that case, the stellar initial mass
function is a direct consequence of the core mass distribution: $M_c
\propto \varepsilon f_p^{-3/2}$.  Conversely, the distribution of
$f_p$ is empirically determined, in such a theory, by the distribution
of stellar (and binary) masses.  For instance,
\cite{1997MNRAS.288..145P} attribute the stellar initial mass function
to the statistical distribution of pressure in simulations of
supersonic, isothermal, unmagnetized turbulence.  We will not adopt
this strategy for the distribution of $f_p$, however, for two
reasons. First, such a theory predicts that the most massive stars
form in the lowest-density regions, which contradicts observations.
Second, cores that create massive stars are significantly more massive
than $M_{\rm BE}$ and therefore highly turbulent
\citep[e.g.,][]{2003ApJ...585..850M}.  We reiterate that we consider
only the typical, low-mass cores, for which thermal support is
naturally significant.

We have not considered static magnetic fields in our treatment of
cores; we comment on this in \S \ref{ObsComp}. 

\subsection{Core Rotational Properties} \label{CoreRot} 
We follow \cite{2000ApJ...543..822B} and \cite{2004ApJ...600..769F} in
assuming that the velocity field is homogeneous and Gaussian random
with a spectrum consistent with equation (\ref{Larson}).  These
authors also assume the components of $\mathbf v$ to be uncorrelated,
reducing the mean vorticity relative to an incompressible velocity
field of the same $\sigma$.  We shall first adopt the assumption of 
uncorrelated velocity components, then consider how the result should
be modified to account for gas pressure. 

Because both $\sigmaNT$ and the specific angular momentum $j$ derive
from $\mathbf v$, one expects $j\propto R_c \sigmaNT(R_c)$, with
statistical fluctuations about this scaling.  This expectation is
borne out in Appendix \ref{Appendix:j}, where we compute the full
distribution of $j$.  Equations (\ref{<j>from<sigma>R}) and
(\ref{jdistribution}) imply, for a velocity field with uncorrelated
components,
\begin{eqnarray} \label{j} 
j &=& 0.34 f_j R_c\, \sigmaNT 
\nonumber \\ &=& 
2.3\times 10^{21} f_j \left(\frac{\Sigmacl}{165\Msun\pc^{-2}}
\right)^{1/2} \left(\frac{R_c}{0.1~\rm pc}\right)^{3/2} {\rm
  \frac{cm^2}s} 
\nonumber \\ &\simeq& 
1.3 \times 10^{21} \frac{f_j}{f_p^{3/2}} T_{10}^{3/2}
\frac{165\Msun\pc^{-2}}{\Sigmacl} {\rm \frac{cm^2}s}
\end{eqnarray} 
where the last line uses $R_c$ from equation
(\ref{coresWithTurb}). 

Since we have assumed $\mathbf v$ to be scale free, $f_j$ is chosen
from a distribution that depends only on the internal structure of the
core, the turbulent spectral slope, and any correlations between
components of $\mathbf v$.  As explained in \S \ref{Appendix:sigma},
$j$ is distributed as a Maxwellian, or, crudely, 
\begin{equation}\label{fj}
\log_{10} f_j = 0_{-0.49}^{+0.16}. 
\end{equation} 

The above results pertain to velocity fields with strictly
uncorrelated components, which is not entirely realistic.  An
uncorrelated flow can be decomposed into an irrotational, potential,
compressive portion and a vortical, incompressive portion.  In Fourier
space, potential flows are aligned with the wavevector and vortical
flows are perpendicular to it. Since there are twice as many degrees
of freedom in the perpendicular direction, an uncorrelated flow
contains two-thirds of its power in vortical motion and one-third in
potential flow.  Since core angular momentum derives entirely from the
vortical part of the flow, it is $\sqrt{3/2}$ times higher in
incompressible turbulence as compared to uncorrelated turbulence of
the same $\sigma$.  Equation (\ref{MachCore}) implies turbulent Mach
numbers of about unity on the core scale, suggesting behavior
intermediate between the incompressible and pressureless limits. It is 
reasonable, therefore, to multiply $f_j$ by $1.11$ (the harmonic mean of
$\sqrt{3/2}$ and unity) to account for correlations introduced by
pressure.  This correction is however rather minor. 

\subsection{Collapse of Cores} \label{Collapse} 
Cores collapse with the accretion rate characteristic of any
self-gravitating body, $\dot{M}= \ceff^3/G$
\citep{1977ApJ...214..488S}.  We assume only a fraction
$\varepsilon$ of the core successfully accretes, so the final stellar mass
is $M = \varepsilon M_c$, and 
\begin{eqnarray}\label{MdotAcc}
\dot{M}_{\rm acc} &\simeq& \varepsilon \frac{\ceff^3}{G}
\nonumber \\ 
&=& \frac{2.8\times 10^{-6}}{f_p^{3/4}} \frac{\varepsilon}{60\%}
     T_{10}^{3/2}~\Msun~\yr^{-1}. 
\end{eqnarray}
There exists an initial phase of more rapid accretion during which the
central region of the core collapses homologously; however,
equation (\ref{MdotAcc}) is valid for most of the mass. 

The theory predicting $\varepsilon\sim 25\%-80\%$
\citep{2000ApJ...545..364M} invokes the removal of material from the
rotational axis by protostellar jets; this somewhat increases the mean
angular momentum of accreting material, but we make no correction for
this effect.

It is important to note that equation (\ref{MdotAcc}) relies on the
assumption that cores are the mass reservoir for stars and collapse in
their entirety.  This accounts neither for the possibility that
accretion might be aborted before the collapse has finished
\cite[e.g.,][]{2001AJ....122..432R}, nor for the possibility that
accretion continues after the core has collapsed
\citep{2001MNRAS.323..785B}.  These effects would alter the amount,
duration, and angular momentum of accreted material; however, we
assume they can be neglected when estimating the characteristic scales
of star formation.

The characteristic disk radius formed during infall is $j^2/(G M)$, or 
\begin{equation} \label{Rdisk}
R_d \simeq \frac{f_j^2}{45 \varepsilon f_p^{1/2}} R_c = 230\  T_{10}\
\frac{f_j^2}{f_p^{3/2}}\frac{60\%}{\varepsilon}
\frac{165\,\Msun\,\pc^{-2}}{\Sigmacl}\ \AU. 
\end{equation}
This is comparable to the critical radius beyond which the disk's
self-gravity can cause local fragmentation, at least in principle; see
equation (\ref{OmegaQProtostellar}) below.  The maximum disk period is
of order 9000 years, but is quite sensitive to the parameters $f_j$,
$\varepsilon$, and $f_p$.

The inverse relation between $R_d$ and $\varepsilon$ results from the
fact that a decrease in $\varepsilon$ does not reduce the specific
angular momentum $j$ of accreted material.  This is true for the
\cite{2000ApJ...545..364M} model, in which an outflow removes material
only along the rotational axis.  In scenarios where $\varepsilon<1$
because accretion is aborted, the disk inherits only low-$j$ material
from the core center and $R_d$ (and the tendency to fragment) are
accordingly suppressed.

\subsection{Comparison to Prior Models}\label{PrevThy}

We have not attempted to model the internal structures of cores under
the influence of both thermal and nonthermal support, as previous
authors have done \citep{1992ApJ...396..631M,MH99,CM99}. Our emphasis
has instead been on the overall mass, radius, and angular momentum
scales of the smallest collapsible objects that can exist in a given
turbulent environment. For this we rely on a Bonnor-Ebert
model, rescaled to account for nonthermal support, and neglect the 
flattening of the density profile this nonthermal pressure should
provide.  Although far from perfect, this approximation is best for
the most thermally-supported cores on which we concentrate.

In our study of core rotation we have followed
\cite{2000ApJ...543..822B} in positing a homogeneous turbulent
velocity field obeying the line width-size relation (\ref{Larson})
that pervades cores and gives them rotation.  Studies of core internal
structure have often assumed instead that the nonthermal pressure is a
function of the local density, e.g., in constructing multi-pressure
polytrope \citep{MH99} or nested polytrope \citep{CM99} core models,
or that it is a function of the local distance from the core's center
\citep{1992ApJ...396..631M}.  Polytropic models are motivated, in
part, by the polytropic behavior of Alfv\'en wave pressure in the WKB
approximation \citep{1995ApJ...440..686M}.  This is an idealization
because of the predominance of long-wavelength perturbations and
because numerical simulations
\citep[e.g.,][]{1998ApJ...508L..99S,1999ApJ...524..169M} indicate that
Alfv\'en waves damp rapidly in molecular cloud conditions.  Our
assumption of homogeneous turbulence is equally idealized: it does not
explain the origin of the turbulent field, and it ignores the
existence of a line width-density relation.  We also have not
accounted for the possibility of a break in the line width-size
relation on scales between the core and its parent clump or cloud.
Nevertheless, our results are in reasonable agreement with
observations (\S \ref{ObsComp}).

Our model is similar in some ways to the TNT (thermal plus nonthermal)
model introduced by \cite{1992ApJ...396..631M} and used recently by
\cite{2003ApJ...585..850M}.  However the TNT model posits a nonthermal
line width that depends on the local radius from the core center: it is
akin to the polytropic models in that it associates turbulence with the
local conditions.  The TNT model also idealizes core density
structures as singular power laws, rather than accounting for the flat
central region that exists prior to collapse: this is significant
because it implies that thermal pressure will always dominate within
some radius.  In contrast, our models argue for comparable
thermal and nothermal contributions even in maximally thermal
cores.  

Our predictions for the mass and radius of precollapse cores,
equations (\ref{coresWithTurb}), are similar to those of the smallest
(thermal) cores in \cite{2003ApJ...585..850M}'s theory.  However these
authors do not account for the possibility of a minimum level of
turbulent support.  As a result, their thermal cores are less massive
and more slowly accreting than we would predict.  Although
\citeauthor{1992ApJ...396..631M} and \citeauthor{2003ApJ...585..850M}
do not address the core angular momentum distribution, their thermal
cores would be significantly more quiescent and more slowly rotating
than the cores modeled here.  

\subsection{Comparison to Observations}\label{ObsComp}

\noindent{\em Masses:}
The mass and radius scales derived in \S \ref{Inits}
(eq.~[\ref{coresWithTurb}]) are consistent 
with the typical properties of prestellar cores observed in nearby
star-forming regions, like Orion B \citep{2001ApJ...559..307J}.  The
core mass spectrum in such surveys typically shows a power law to
masses above the minimum thermally-supported mass identified in this
section.  

\noindent{\em Nonthermal linewidths:}
\cite{1999ApJS..125..161J} classified presetellar cores according to
their association with cores and {\it IRAS} sources.  In their
database, those cores that are truly starless in this comparison have
median ratios of nonthermal to thermal linewidths about 35\% smaller
than suggested by equation (\ref{MachCore}) -- see their figure 4.
The observed distribution overlaps our prediction.  Some of the
discrepancy may be attributable to rotational motions that do not
contribute to the (beam-resolved) linewidth, or to static magnetic
fields.

\citeauthor{1999ApJS..125..161J} observe a correlation between the
thermal and nonthermal linewidths: $\sigmaNT\propto \sigmath^{3.83}$
(their figure 7 and table B7).  At the cold and quiesent end of this
correlation are L1498 and L1517B, studied in detail by
\cite{2004A&A...416..191T}.  This trend is not explained in our
theory, in which equation (\ref{MachCore}) gives $\sigmaNT \simeq
\sigmath$. 

Recall however that equation (\ref{MachCore}) only represents the
characteristic scales of cores.  It relies on several assumptions:
1. that the core is at the brink of collapse; 2. that it is free of
magnetic support; and 3. that it has the minimum level of turbulence
consistent with assumptions 1 and 2 in the context of a homogeneous
model for the turbulent velocity field.   An individual core need not 
adhere to these assumptions, although we have argued that they suffice
to set characteristic scales.   In the case of the very quiescent
cores\footnote{These are not characteristic of the
172 \cite{1999ApJS..125..161J} cores: for instance, L1498 is their
smallest core.}  L1498 and L1517B, we suspect that magnetic fields
rather than turbulence are supplying a significant amount of
nonthermal support. 

\noindent{\em Nonthermal support:}
\cite{2001ApJ...559..307J} identify an effective temperature, using
Bonnor-Ebert, models which is typically 20-40 K.  Many of these
clumps' actual temperatures measured at less than 20 K, with 10 K
being typical.  A significant contribution of nonthermal pressure --
as in our equation (\ref{MachCore}) -- offers an explanation for this
discrepancy.

It is interesting to note that this level of nonthermal support holds
even in the very quiescent cores L1498 and L1571B, for which
\cite{2004A&A...416..191T} find $T_{\rm eff}=$ 28 and 20 K,
respectively, in their Bonnor-Ebert fits.  Compared to the measured
thermal temperatures $T_c =$ 10 and 9.5 K, we note that the ratio is
consistent with equation (\ref{MachCore}) although the linewidths are
not.  Although part of the nonthermal support of these cores is
rotational, most is likely magnetic.

\noindent{\em Rotation rates:}
As for core rotation rates, equation (\ref{j}) agrees with the value
$j\sim 10^{21} {\rm ~cm^2\;s^{-1}}$ deduced by
\cite{1993ApJ...406..528G} for cores of similar mass.  The second line
differs from \cite{2000ApJ...543..822B}'s formula $j=7\times 10^{20}
(R_c/0.1\pc)^{3/2} {\rm cm^2\; s^{-1} }$.  However,
\citeauthor{2000ApJ...543..822B}'s results were numerical; our own
unpublished numerical experiments, which were based on their method,
exhibit a bias to underestimate $j$ by about the same factor.
We believe this arises due to the periodic boundary conditions
employed in the numerical grid; due to limited dynamical range, these
impose correlations on the box scale that affect the spectrum as
sampled on the core scale and smaller.

The rotation rates observed by \cite{2004A&A...416..191T} for the very
quiescent cores L1498 and L1517B are only a factor of a few smaller
than the typical values predicted by equation (\ref{j}), and are 
within the distribution described by equation (\ref{fj}). 

\noindent{\em Accretion rates:}
\cite{2003MNRAS.344..461M} review a range of observational evidence
that mass accretion can significantly exceed the characteristic value
$c_s^3/G$ in an early (class 0) accretion phase.  They suggest the
initial phase of homologous collapse, or collapse induced by an
external impulse \citep[see also][]{2003MNRAS.340..870H}, as possible
explanations.  These possibilities are both very reasonable, but note
that we predict an accretion rate of roughly $2.6 c_s^3/G$, on account
of the nonthermal support (eq.~[\ref{MdotAcc}]).  Therefore, the need
to invoke alternative explanations for the observed protostellar
accretion rates is partially alleviated.

\noindent{\em Magnetic fields:}
Finally, we comment on the neglect of static magnetic fields in our
calculations.  Employing the Chandrasekhar-Fermi method,
\cite{2004ApJ...600..279C} find that static magnetic fields in
prestellar cores could either be subdominant or comparable in
magnitude with thermal pressure, depending on the application of a
geometrical correction factor to their observations.  Theoretically,
one expects that ambipolar diffusion will sap the influence of
magnetic fields prior to core collapse.  In addition, 
inferences of short core lifetimes \citep[e.g.,][]{2002AJ....124.2756V}
and observations of high core column densities \citep{1998ApJ...494..587N}
are taken to indicate that static magnetic fields are not significant
in core support.
On theoretical grounds, \cite{1983ApJ...270..511Z} find that
ambipolar diffusion is accelerated in turbulent gas.  
We conclude that our approximation of zero mean field is adequate
given current uncertainties, while recognizing that this should be
re-examined in future work.  
As discussed above, magnetic fields may be especially important for
understanding the thermal-nonthermal linewidth correlation seen by
\cite{1999ApJS..125..161J} and the nonthermal support of quiescent
cores seen by \cite{2004A&A...416..191T}.

\section{Prompt Fragmentation}\label{Impulsive} 
The first opportunity for fragmentation is in the initial phase of
collapse, which follows from the homologous contraction of the core's
central region.  In this phase, the mass accretion rate can exceed equation
(\ref{MdotAcc}).  If the core's central region rotates sufficiently
rapidly, it will form a disk that can fragment.  

This phase of fragmentation has been studied in detail recently by
\cite{2003ApJ...595..913M}, who consider the initial collapse of slowly
rotating Bonnor-Ebert spheres.  \citeauthor{2003ApJ...595..913M} find
that fragmentation occurs by one mode or another so long as $\Omega_c
t_{\rm ff,ctr}\gtrsim 0.045$, where $\Omega_c$ is the rotation rate
and $t_{\rm ff,ctr}$ is the free-fall time evaluated at the middle of
the initial state.  Using equations (\ref{coresWithTurb}), (\ref{j}),
and the moment of inertia $I_c=0.2827 M_c R_c^2$, we find
\begin{equation}\label{MatusmotoHanawaParam}
\Omega_c  t_{\rm ff,ctr} \simeq 0.25 \frac{f_j}{f_p^{1/4}}. 
\end{equation} 
By this criterion, then, all but the most slowly rotating cores
($f_j<0.18 f_p^{1/4}$; about $3\%$ of the population according to
eq. [\ref{jdistribution}]) should undergo fragmentation in the initial
collapse phase. 

However, this conclusion is not robust.  Note, first, that
\cite{2003ApJ...595..913M} rely on the approximation of barotropic
core collapse (isothermal, for $n_H<10^{13}$ cm$^{-3}$) rather than a
detailed thermodynamic calculation.  
\cite{2000ApJ...528..325B} caution that protostellar collapse simulations
employing the Eddington approximation differ significantly with regard
to fragmentation from those that use a barotropic
law.  Recently, \cite{2004astro.ph.12357L} has argued that thermal evolution
plays a critical role in determining stellar masses. 
Likewise, numerical and analytical studies of disk fragmentation
\citep{2001ApJ...553..174G,2003ApJ...590.1060P,2004astro.ph..6469R,
2003astro.ph..7084L} stress the importance of cooling.  We will return
to this issue in detail in \S \ref{Instab} and add some comments in \S
\ref{NumericalImplications}. 

Second, \citeauthor{2003ApJ...595..913M}'s calculations stop after at
most $0.07 \Msun$ of several $\Msun$ have accreted into fragments.  As
they point out, the future evolution of these fragments is difficult
to predict.  
We note that the remaining mass accretes through a disk around the
newly formed objects.  The specific angular momentum of this material
greatly exceeds that of the observed fragments.  As shown by
\cite{1997MNRAS.285...33B}, a binary tightens due to gravitational
interaction with the circumbinary disc.  We suspect that this drives
prompt protostellar fragments to merge.  Future simulations will be
required to test this hypothesis, especially for the complicated case
of multiple prompt fragmentations.  If prompt fragmentation {\em does}
succeed, it should produce tight binaries with $j$ significantly below
the typical value cited in equation (\ref{j}) -- an therefore,
significantly below the median of the binary distribution
\citep{1991A&A...248..485D}.

For both of these reasons we take equation
(\ref{MatusmotoHanawaParam}) to be only {\em suggestive} of
fragmentation in the initial, impulsive collapse phase.  We probe
fragmentation in later stages of accretion in the upcoming sections.

\section{Steady Protostellar Disks: Local Gravitational Instability
  and Local Fragmentation}
\label{Instab} 

We now explore the stability of the steady protostellar accretion
disks that form during the collapse of cores like those described in
\S \ref{Inits}, during the phase of main accretion during which the
accretion rate is given by (\ref{MdotAcc}).  There are two modes by
which self-gravitation can trigger disk fragmentation, one of which is
subject to a local instability criterion \citep{1964ApJ...139.1217T}
and the other a global one \citep{1990ApJ...358..495S}.  We focus
initially on the local instability and return to the global one in \S
\ref{global}.  The consequences of Toomre's instability have been
greatly elucidated by the numerical simulations of
\cite{2001ApJ...553..174G}, as described below.

\cite{1990ApJ...358..515L} have carried out an analysis of
protostellar disk evolution, based on initial conditions qualitatively
similar to those sketched in \S \ref{Inits}.  They implemented a
gravitational viscosity that rose as Toomre's instability parameter,
\begin{equation}\label{def:Q}
Q\equiv {c_d\Omega\over \pi G\Sigma},
\label{Q}
\end{equation}
declined through its critical value of unity.  Here $\Omega$ and
$\Sigma$ are the disk's orbital frequency and column density,
respectively; and \[c_d\equiv \sqrt{\partial \int_{-\infty}^{~\infty}{
P\, dz} \over \partial \Sigma }\] (where $z$ is altitude) is the
signal speed in the disk, which is slightly above the midplane
isothermal sound speed $c_s \equiv \sqrt{P/\rho}$.  
With this prescription for angular momentum transport, they found $Q$
saturating typically in the range 0.2-0.5.  This study lacked,
however, two physical important physical effects: 1. the nonlocal
nature of gravitational angular momentum transport when $Q>1$; and
2. the dissolution into fragments of disks with $Q\lesssim 1$.  Both
shortcomings are addressed by the nonlinear simulations of
\cite{1998ApJ...504..945L} and \cite{2001ApJ...553..174G}:

1. When gravitational instability is weak and $Q>1$, spiral modes
saturate at low amplitudes via mode-mode coupling
\citep{1998ApJ...504..945L}; in this state, redistribution of angular
momentum is nonlocal and is described only approximately by a local
prescription, i.e., by \citeauthor{1973A&A....24..337S}'s $\alpha$
parameter.  On the other hand, strongly unstable disks enter a state
of ``gravitoturbulence'' \citep{2001ApJ...553..174G} in which the
coherence length is only a few effective scale heights
\[H\equiv \frac{c_s}{\Omega}.\]  
In this state angular momentum transport is a local process \citep[see
also][]{2004MNRAS.351..630L}.  At the brink of fragmentation,
\citeauthor{1973A&A....24..337S}'s viscosity parameter $\alpha$ takes
maximal value, which \cite{2001ApJ...553..174G} shows to
be\footnote{We define $\alpha$ using the isothermal sound speed $c_s$,
rather than the disk signal speed $c_d$, in equation
(\ref{mdotVisc}).  As a result our $\alpha$ differs by a factor $\gamma_{2d}$
relative to its definition in \cite{2001ApJ...553..174G}.}
\begin{equation}\label{alphamax}
\alpha = \frac{4}{9 (\gamma_{2d}-1) \Omega t_{\rm cool}}
\end{equation} 
if $t_{\rm cool}$ is the cooling timescale in the critical state. 

We recall that $\alpha$ is related to the mass accretion rate through
the disk:
\begin{equation}\label{mdotVisc}
\dot{M}_{\rm visc}= {3\pi\alpha\Sigma c_s^2\over \Omega}. 
\end{equation}
The subscript ``visc'' is added to distinguish the accretion rate due
to local viscous transport from, for instance, the rate of mass supply
(eq. [\ref{MdotAcc}]).  

\cite{2003MNRAS.339..937G} points out that
angular momentum loss in magnetized winds can lead mass to accrete
more rapidly than equation (\ref{alphamax}) predicts.  We however assume that
strong winds do not develop over most of the disk radius, and we adopt
equation (\ref{alphamax}) as a maximal value.

2. By identifying the precise boundary between gravitoturbulent and
fragmenting disks, \cite{2001ApJ...553..174G} provides the means to
discriminate stable mass accretion from the formation of new bound
objects.  A number of papers argue that the precise outcome of
fragmentation depends sensitively on the thermal properties of the
disk \citep[e.g.,][]{2003ApJ...590.1060P}; nevertheless the threshold
for instability is given adequately by equation (\ref{brink}).

These insights reveal two points about
\citeauthor{1990ApJ...358..515L}'s simulations.  First, they estimate
in a reasonable fashion the dynamics of protostellar disks in their
main accretion phase (since their estimate of $\alpha$ approximates
\citeauthor{2001ApJ...553..174G}'s result when $Q=1$); and second,
that these disks are close enough to $Q=1$ to merit an careful
investigation of whether fragmentation does indeed occur.

The results of Gammie's two-dimensional simulations
\citep[corroborated in three dimensions by][]{2003MNRAS.339.1025R}
provide a practical, local description for the disks that have entered
a state of self-gravitating turbulence.  This state separates disks
that are relatively smooth from those that have fragmented into swarms
of clumps; it also is a state in which the mean midplane conditions
(density, temperature, etc.) are reasonably constrained.  We use it
as a litmus test for fragmentation in a variety of circumstances.

\subsection{Local criteria for fragmentation by Toomre's
  instability}\label{localcriteria}  

\cite{1964ApJ...139.1217T} demonstrated the onset of axisymmetric
instability in disks with $Q<1$; although other modes grow for
somewhat higher $Q$, \cite{2001ApJ...553..174G} verified that $Q\simeq
1$ at the boundary of fragmentation. 

To assess whether the disk can process accretion without fragmenting,
we adopt 
\begin{equation}\label{cd-cs}
c_d^2 = \gamma_{2d} c_s^2, ~~~{\rm
  where}~~~\gamma_{2d}\equiv { \partial\ln
  \int_{-\infty}^{\,\infty}{ P\, dz} \over \partial \ln\Sigma}
\end{equation}
in equations (\ref{Q}) and (\ref{alphamax}).  In the limits of weak
self-gravity  \citep{1986MNRAS.221..339G} or strong self-gravity
\citep{2001ApJ...553..174G}, repsectively, 
\begin{equation}\label{gamma2d-vs-gamma} 
\gamma_{2d} =\left\{\begin{array}{lc} 
  \frac{3\gamma-1}{\gamma+1} & (Q\gg 1) \\
  3-{2\over\gamma} & (Q\ll 1) \end{array}\right.
\end{equation}   
where $\gamma$ is the adiabatic index of disk gas.

\cite{2001ApJ...553..174G} adopted the specific value $\gamma_{2d}=2$,
corresponding to $\gamma=3$ or 2 for strong or weak self-gravity,
respectively.  Either value is stiffer than a realistic disk.  In the
temperature range of interest, a protostellar disk obeys $\gamma=5/3$
because rotational degrees of freedom are not excited;\footnote{Shock
heating and radiative transfer alter the effective $\gamma$ from its
adiabatic value, especially for disks close to fragmentation.  However
it is the adiabatic $\gamma$ that enters in the theory of
\cite{2001ApJ...553..174G}.}  then $\gamma_{2d} = 3/2$ or $9/5$ in the
strong and weak self-gravity limits, respectively.  At the onset of
fragmentation $Q\simeq1$, and we assume that
\[\gamma_{2d}\simeq 1.65,~~~~~~~({Q=1})\] an intermediate value.  We
also assume that the critical cooling time remains roughly one half orbital
period, i.e., $\Omega t_{\rm cool}=3$, for the softer equation of
state; this remains to be checked by future numerical experiments.
With these choices, equation (\ref{alphamax}) gives
\begin{equation}\label{brink}
\alpha = 0.23. ~~~~~~~({Q=1})
\end{equation}
Weak and strong self-gravity give $\alpha=0.30$ and
0.19, respectively.  We adopt $\alpha=0.23$ as a fiducial maximal value,
but consider this uncertain by about 50$\%$. 

The condition $Q=1$ may immediately be expressed in terms of $c_s$, 
the midplane density $\rho$, and the midplane pressure $p$: 
\begin{equation}\label{rho} 
c_s = \frac{\pi  G \Sigma}{\gamma_{2d}^{1/2}\Omega}, ~~
\rho={{\gamma_{2d}}^{1/2} \Omega^2\over 2\pi G},~~{\rm and}~~
p = {\pi\over 2{\gamma_{2d}}^{1/2}} G \Sigma^2 
\end{equation}
where the middle expression uses the approximate relation $\rho =
\Sigma\Omega/(2 c_s)$, and the last expression uses $p=\rho c_s^2$.
These relations are only approximate, as they neglect the disk's
self-gravity 
in its vertical structure,
but they suffice to delineate the onset of instability. 

An important additional criterion comes from combining eqs. (\ref{MdotAcc}),
(\ref{Q}), and (\ref{mdotVisc}) to give 
\begin{equation} \label{MdotViscOnMdotAcc}
\frac{\dot{M}_{\rm visc} }{\dot{M}_{\rm acc}} = 
\frac{3\alpha \gamma_{2d}^{1/2}}{\varepsilon Q}
  \frac{c_s^{3}}{\ceff^{3}} . 
\end{equation} 
Conversely, if the disk
processes gas in steady state ($\dot{M}_{\rm visc} = \dot{M}_{\rm
acc}$) then
\begin{equation}\label{Qsteady} Q = \frac{3\alpha
    \gamma_{2d}^{1/2}}{\varepsilon} 
\frac{c_s^3}{\ceff^{3}}. ~~~~~~ (\rm steady-state)
\end{equation} 
Using $\gamma_{2d}=1.65$ and $\alpha=0.23$ to describe the onset of
fragmentation, we find that the disk can process steady accretion with
$Q>1$ only when  
\begin{equation} \label{Cs>Ceff}
c_s >  1.04 \varepsilon^{1/3}\ceff;   
\end{equation}
i.e., using eq. (\ref{MachCore}) to describe a typical unstable core,  
\begin{equation}\label{T>Teff}
T>1.08 \varepsilon^{2/3}T_{\rm eff} \simeq \frac{1.56}{f_p^{1/2}}
\left(\frac{\varepsilon}{60\%}\right)^{2/3} T_c 
\end{equation} (where $T$ refers to the
midplane temperature); otherwise fragmentation will occur where the
gas falls.  For typical core temperatures of order 10 K, disks colder
than about 15.6 K are unable to process the high accretion rate and
will fragment into clumps.  This critical temperature is rather
insensitive to the uncertainties described after equation (\ref{brink}): it
varies as $\alpha^{-2/3} \gamma_{2d}^{-1/3}$.

\subsubsection{Self-luminous disks} \label{self-luminous}

If the heat generated locally by viscosity is responsible for the
midplane temperature, then criterion (\ref{Cs>Ceff}) may be evaluated
directly.  Our treatment follows \cite{1990ApJ...358..515L} and
\cite{2003astro.ph..7084L}, among others.  In thermal steady state,
the flux of viscous energy radiated by each face of the disk is
related to the accretion rate through
\begin{equation} \label{FluxVisc}
F_v={ 3\Omega^2 \dot{M}\over 8\pi}.
\end{equation}
The flux can also be derived from radiation transfer across an optical
depth $\kappa \Sigma/2$ from the disk midplane to its surface;
therefore 
\begin{equation} \label{FluxRad}
F_r = \sigma T^4 \times \left\{ 
\begin{array}{cc} 
  4 \tau_{\rm Pl} & (\tau \ll 1)  \\ 
  {8}/({3\tau_{\rm R}}),& (\tau \gg 1)  \\ 
\end{array} 
\right.
\end{equation}
where $\tau_{\rm R,Pl}= \kappa_{\rm R, Pl}\Sigma/2$ is the optical
depth 
corresponding to Rosseland or Planck opacity, $\kappa_{\rm R}$
or $\kappa_{\rm Pl}$, respectively.\footnote{
Recently, \cite{2003ApJ...597..131J} have published numerical
experiments on razor-thin discs and have shown that
Eq.~(\ref{FluxRad}) is generally inaccurate when the disc develops
strong gravitationally-driven turbulence. However, in their model the
breakdown of Eq.~(\ref{FluxRad}) is serious when the opacity is a
sensitive function of temperature, which occurs only when the dust
begins to evaporate at above $1000$ K. Temperatures this high are
clearly irrelevant for protostellar discs beyond a fraction of an AU,
and thus for the problem at hand Eq.~(\ref{FluxRad}) is adequate.  We
also note that when the opacity is temperature sensitive, then the
one-zone model of \citeauthor{2003ApJ...597..131J} does not treat
faithfully the vertical radiative transport in a real disc.}  The
factor $16/3$ arises in the optically thick case when the dissipation
rate per unit mass is a constant \citep[e.g.,][]{1997ApJ...477..398C}
and is only approximate.  For the disk to achieve an equilibrium
temperature in which local heating balances radiative cooling, we must
have $F_v=F_r$.

The midplane temperature and sound speed are related via
\begin{equation}\label{cs}
c_s^2={k_B T\over\mu }
\end{equation}
where $\mu\simeq 2.3 m_p$ is the mean molecular weight. We neglect the
radiation pressure in this expression.  This is entirely
justified for present-day star formation, since when $Q=1$
the ratio of gas to radiation pressure is
\[ 
\frac{3k_B \Omega^2}{2 \pi G a \mu T^3} = 11 \left(\frac{1000\
  \yr}{\rm period}\right)^2 
\left(\frac{500~\rm K}{T}\right)^3 \] using eq. (\ref{rho}).

For the opacity we adopt
\begin{equation} \label{kappa} 
\left(
\begin{array}{c} 
 \kappa_{\rm R} \\ 
\kappa_{\rm Pl} \\ 
\end{array} \right) =  \kappa_0 T^2 = \left( 
\begin{array}{c} 
   3.0 f_{\kappa \rm R} \\ 
   8.4 f_{\kappa\rm Pl} \\ 
\end{array} 
\right)
\times  10^{-4} T^2 {\rm ~cm^2~g^{-1} }. 
\end{equation}
This describes the ice branch of \cite{2003A&A...410..611S}'s ice
grain opacities in their ``composite aggregate'' model;
eq. (\ref{kappa}) fits their results around 16 K.  Note that
\cite{1994ApJ...437..879A} computed somewhat lower opacities
\citep[$f_{\kappa \rm R} \simeq 0.67$; see
also][]{1994ApJ...427..987B}.  \cite{1996A&A...311..291H} also derive
a lower value ($f_{\kappa\rm R} \simeq 0.43$) after allowing for dust
agglomeration in the parent molecular core.  We include the two
$f_\kappa$ factors to account for variations in metallicity, dust
processing, and uncertainties in opacity modeling.

Equations (\ref{FluxVisc}), (\ref{FluxRad}), (\ref{cs}), and
(\ref{kappa}) provide enough information to solve $F_v = F_r$ for
$T(r)$, given an accretion rate $\dot{M}$.  As discussed above, the
local accretion rate (eq. [\ref{mdotVisc}]) must match the supply
(eq. [\ref{MdotAcc}]).  To determine the onset of fragmentation, we
also impose $Q\rightarrow 1$ (eq. [\ref{Q}]).  We find that
fragmentation ensues when
\begin{eqnarray} \label{OmegaCritProtostellar} 
\Omega = \Omega_{\rm cr}  \times \left\{ 
\begin{array}{cc} 
\left(\frac{\ceff}{c_{\tau}}\right)^{10},
 & c_s < c_{s,\tau} ~ (\tau \ll 1)  \\ 
1, & c_s > c_{s,\tau} ~ (\tau \gg 1)  \\ 
\end{array} 
\right.
\end{eqnarray} 
where 
\begin{eqnarray} \label{OmegaQProtostellar}
\Omega_{\rm cr} &=& 3.60 \left[\frac{G^2 \mu^2 \sigma }{\alpha
    \kappa_0 \gamma_{2d} k_B^2}\right]^{1/3} 
\nonumber \\ &=&  
\frac{4.3 \times 10^{-10}}{ f_{\kappa\rm R}^{1/3} }
\left(\frac{\mu}{2.3 m_p}\right)^{2/3} 
\left(\frac{0.23}{\alpha} 
\frac{1.65}{\gamma_{2d}}\right)^{1/3} 
 {\rm s}^{-1},\nonumber\\
\end{eqnarray} 
corresponding to a period of 460 years for the fiducial parameters
(400 years for \citeauthor{1994ApJ...437..879A}'s opacity; 350 years
for \citeauthor{1996A&A...311..291H}'s opacity), and
\begin{eqnarray} \label{ceffTauProtostellar}
c_{s,\tau} &=&  \frac{0.174}{f_{\kappa\rm R}^{1/30}f_{\kappa\rm Pl}^{1/10}} 
\left(\frac{\alpha}{0.23}\right)^{1/15}  
\left(\frac{1.65}{\gamma_{2d}}\right)^{1/30} \nonumber\\ &~&\times 
\left(\frac{2.3 m_p}{\mu}\right)^{8/15} ~{\rm km~s^{-1}}.  
\end{eqnarray} 

To summarize, disks with opacity law (\ref{kappa}) that are warmed
solely by their own viscosity will be subject to local fragmentation
in any region whose orbital frequency is less than the value given in
eq. (\ref{OmegaCritProtostellar}).

It is worthwhile to emphasize two points about the above equations. 
First, $\Omega_{\rm cr}$ is an upper bound for the frequency of a
locally unstable disk, i.e., all locally unstable disks have periods
of roughly 460 years or longer.  This bound depends {\em only} on the
opacity law of the disk midplane and is independent of the mass
supply rate and disk surface density (so long as the disk is optically
thick).  This result is specific to a cancellation that occurs when
the opacity varies as $T^2$. 

Second, the lower bound on the period at which fragmentation can occur
is pushed to even higher values if the disk is optically thin or if it
is affected by external irradiation. 
The division between optically thin and thick fragmentation depends
only on the midplane sound speed $c_s$; the critical value
$c_{s,\tau}$ corresponds to a midplane temperature of roughly
8.4 K.  But recall that the fragmentation temperature is determined by
the mass supply rate to be roughly $T\simeq 16$ K for cores with
$T_c=10$ K
(eq. [\ref{T>Teff}]).  This implies that disks are typically opticall
thick at the fragmentation bound Indeed, using equations
(\ref{T>Teff}) and (\ref{ceffTauProtostellar}), all cores with
\begin{equation}\label{TcMakesThickFrag}
T_c > 5.4 \frac{f_p^{1/2}}{f_{\kappa\rm Pl}^{1/5} f_{\kappa\rm
 R}^{1/15}}
 \left(\frac{60\%}{\varepsilon}\right)^{2/3} {\rm K}
\end{equation}
will produce disks that are significantly more prone to fragmentation
than those from colder cores because of the transition to optically
thick disks. (In equation [\ref{TcMakesThickFrag}], we suppress the
variation $T_c \propto \alpha^{4/5} \gamma_{2d}^{4/15}\mu^{-1/15}$
around fiducial values of these parameters.)

Equation (\ref{OmegaCritProtostellar}) argues strongly against the
formation of giant planets by direct disk fragmentation, in agreement
with the recent work by \cite{2004astro.ph..6469R}.  This argument
gains strength when the effect of stellar irradiation is considered. 

{\em Fragment masses.} We pause here to note the characteristic masses
of fragments formed by local gravitational instability.  This topic is
investigated in detail by \cite{2003astro.ph..7084L} and
\cite{2004ApJ...608..108G}.  If fragments form, their initial mass is
of order $\Sigma (2\pi H)^2$, i.e., roughly Jupiter's mass for both
the optically thick and optically thin cases if $T_c=10$ K (but
scaling as $T_c^{3/2}$ and $T_c^{-7/2}$ respectively).  Continued
growth is likely up to the gap-opening mass, which we estimate at 36
Jupiter masses for fiducial parameters, and perhaps beyond.   Local
fragmentation is therefore most relevant to the origin of planetary
\citep{1997Sci...276.1836B} or brown dwarf \citep{2002MNRAS.332L..65B}
companions.

\subsection{Irradiation}\label{irrad}

The preceding section identified a critical midplane temperature for
local fragmentation (typically $\sim 15.6$ K in disks around low-mass
protostars, eq. [\ref{T>Teff}]), and a critical disk period
(typ. $\sim 460$ yr) above which viscous heating permits cooler
midplane temperatures.  However as we mentioned before, the argument
in the preceding section neglects external irradiation of the disc.
Here we self-consistently take irradiation into account and show that
it completely prevents local fragmentation for a wide range of
core parameters.

To illustrate its importance, suppose irradiation normal to the disk
surface is a fraction $\fF$ of the spherical flux.  The equilibrium
temperature of an optically thick disk is then given by $\sigma T^4 =
\fF L/(4\pi R_d^2)$, if $L$ is the accretion luminosity; or
\begin{equation}\label{irradThick}
94 \fF^{1/4} \left(\frac{T_\star}{4500}\right)^{1/3}
\left(\frac{460\,\yr}{\rm period}\right)^{1/3} 
\left(\frac{\dot{M}_{\rm acc}}{2\times
  10^{-6}\Msun\,\yr^{-1}}\right)^{1/6} {\rm K} 
\end{equation}
where the stellar effective temperature is $T_\star$.  (Note this is
independent of the stellar mass $M$, if referenced to orbital period.)  
If the disk is optically thin, its temperature is generically higher
than given above: for a local radiation energy density is $a T_{\rm
  egy}^4$ and color temperature is $T_c$, the equilibrium temperature $T$
of an optically thin region is 
\begin{equation}\label{irradThin}
T^4 \kappa_{\rm Pl}(T)  = T_{\rm egy}^4 \kappa_{\rm Pl}(T_c) 
\end{equation} 
where $\kappa_{\rm Pl}$ refers to the Planck average opacity.  Since
$\kappa_{\rm Pl}(T)$ is monotonic in the temperature range of
interest,
$T_{\rm egy}<T<T_c$.  Furthermore, since the flux at the
disk surface cannot exceed $a T_{\rm egy}^4 c$, an optically thin disk
must be warmer than a thick disk in the same radiation
environment.\footnote{The lower limit on $T$ in the thin case is
$\sqrt{2}$ below that in eq. (\ref{irradThick}), but this factor is
not significant.}

The flux suppression factor $\fF$ can be quite small in the case that
the disk is illuminated at a grazing incidence from the stellar
photosphere \citep[e.g.,][]{1987ApJ...323..714K,1997ApJ...490..368C}.
For an accreting protostar, however, reradiation from the stellar
envelope tends to illuminate and heat the disk
\citep{1990ApJ...355..635K,1993ApJ...412..761N,1996ApJ...461..956C,
1997ApJ...477..398C,1997ApJ...474..397D}. 

To gauge this, consider the column of the infall envelope.  
The spherically averaged column density outward from some radius $r$ is 
$\Sigsph(r) = \dot{M}/[2\pi r v_{\rm esc}(r)]$.
Evaluated at the centrifugal radius in eq. (\ref{Rdisk}),
\begin{equation}\label{SigmaSpherical-at-Rd}
\Sigsph(R_d) = 0.43 \frac{\varepsilon}{60\%} 
 \frac{f_p^{3/4}}{f_j} \Sigmacl. 
\end{equation} 
In spherical infall $\Sigsph(r)$ diverges inward as
$r^{-1/2}$.  For rotating infall, however, only low-$j$ material
persists within $R_d$; this makes $\Sigsph(R_d)$ a
characteristic value for the column {\em inward} from $R_d$, as well.

Intriguingly, the characteristic column $\Sigsph(R_d)$ is roughly
$\Sigmacl/2$, i.e., half that of the parent clump or cloud, at the end
of accretion.  (Beforehand it is much greater: the column to the disk
edge varies as $t^{-2}$ during steady accretion.)
Deviations from spherical symmetry due to angular momentum are given
by the rotating-infall solution of \cite{1984ApJ...286..529T}.
Additionally, the protostellar jet will scour the axis to produce a
cavity. 

Since star formation appears to be limited to regions of high visual
extinction \citep[][as suggested theoretically by
\citeauthor{1989ApJ...345..782M} 1989]{1998ApJ...502..296O,
  1999PASJ...51..895H, 2004ApJ...611L..45J},  and since
$\Sigsph(R_d)\simeq \Sigmacl/2$, 
starlight is absorbed and reprocessed after it traverses a fraction of
the infall column.  This reprocessing is strongly affected by the
presence or absence of an outflow cavity: therefore, we consider both
cases below.

The temperature of an irradiated disk depends on the optical depth
of the disk to the incoming radiation and to its own thermal
emission.  For a disk at the fragmentation threshold with $Q=1$, 
\begin{eqnarray}\label{SigmaDisk}
\frac{\Sigma(R_d)}{\Sigsph} &=& \frac{3\alpha}{2^{3/2}}
\left(\frac{H}{r}\right)^2 \nonumber \\ &=& 
210 \frac{f_p^{1/2}}{f_j^2}\left(\frac{\varepsilon}{60\%}
\right)^{4/3}. 
\end{eqnarray} 
The first line derives from equations (\ref{def:Q}), (\ref{cd-cs}),
(\ref{Cs>Ceff}), and (\ref{SigmaSpherical-at-Rd}); the second line
applies the results of \S \ref{Inits} (and varies as
$(\gamma_{2d}/\alpha)^{1/3}$ around our fiducial values).  The outer
disk column is thus of order 90 $\Sigmacl$, or $\sim 3$ g cm$^{-2}$
for typical clouds.  According to the opacity law (\ref{kappa}), the
disk will be vertically optically thick to Planck radiation of
color temperature exceeding  \[19.6 \frac{f_j^{3/2}}{f_{\kappa \rm
    Pl}^{1/2} f_p^{5/8} }
\left(\frac{165\Msun\pc^{-2}}{\Sigmacl}\right)^{1/2} {\rm K}. \]
The disk's opacity to radiation from dust of a certain temperature is
higher than to a Planck field of that color temperature: $f_{\kappa\rm
  Pl}\simeq 2.5$ for dust radiation.  This lowers the critical
illuminating temperature to about 12 K, and oblique illumination lowers it   
further still.  The disk can therefore be assumed optically thick to
irradiation, although it is only marginally thick with respect to its own
radiation at a fragmentation temperature of $\sim 15.6$ K.

\subsubsection{No outflow cavity} \label{nocavity}

The reprocessing radius typically lies well inside $R_d$ but outside
the dust evaporation radius, because the latter is optically thin in
the end phases of accretion.  Disk irradiation is therefore due
primarily to the second reprocessing of starlight at radii around
$R_d$.  We shall identify the dust temperature at the reprocessing 
surface, which affects the amount of re-absorption and re-emission
above the disk. 

Close to the protostar, the angle-averaged column density in the
\cite{1984ApJ...286..529T} infall profile is $\bar{\Sigma}(r) =
1.17(r/R_d)^{1/2} \Sigsph(R_d) + {\cal O}(r/R_d)^{3/2}$.  We identify
the reprocessing radius $r_1$ with the surface of optical depth unity,
then $\bar{\Sigma}(r_1) \kappa_\star = 1$, where $\kappa_\star\equiv
\kappa_{\rm Pl}(T_\star)$ is the opacity to starlight of color
temperature $T_\star$.  In addition, equation (\ref{irradThin})
determines the dust temperature $T_1$ at this location: $\kappa_{\rm
Pl}(T_1) T_1^4 = \kappa_\star L/(4 \pi r_1^2 a c)$.  Combining these,
\begin{equation}\label{Reprocessing} 
\frac{\kappa_{\rm Pl}(T_1)}{\kappa_\star^5} \sigma T_1^4 = \frac{L
\Sigsph^4}{26.7}. 
\end{equation} 
The left hand side is a somewhat complicated function of $T_1$ because
the dust composition, a function of $T_1$, determines $\kappa_\star$
and $\kappa_{\rm Pl}(T_1)$; metallicity and the stellar temperature
$T_\star$ also enter.\footnote{$T_1$ also depends weakly on the gas
density at $r_1$, which affects the evaporation temperatures of the
dust components; this is ignored in eq. \ref{Reprocessing}.}  The
right hand side depends on model parameters.  A self-consistent
solution of (\ref{Reprocessing}) is not always available; this signals
a change in the dust composition and fixes $T_1$ at a composition
boundary.  Nevertheless, this equation demonstrates that the
reprocessing temperature $T_1$ is a pure function of $L\Sigma_{\rm
sph}^4$ for fixed $T_\star$ and metallicity, so long as the infall is
optically thick to starlight.  In a range of $T_1$ for which the dust
composition is constant, $T_1\propto L^{1/6}\Sigsph^{2/3}$
because $\kappa_{\rm Pl}(T_1)\propto T_1^2$. 

\begin{figure*}
\centerline{\epsfig{figure=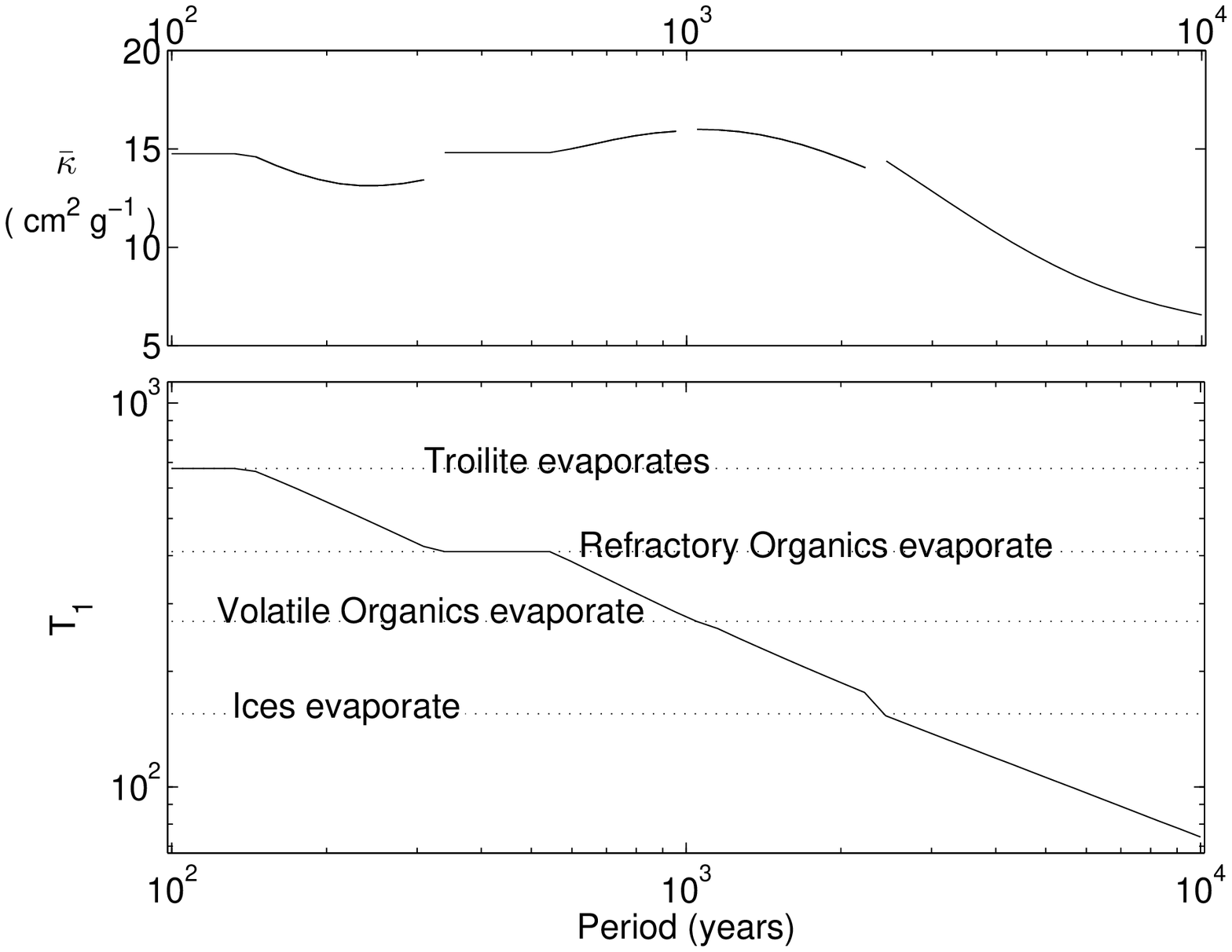,width=3.5in}} 
\caption[] {\footnotesize Lower panel: Temperature $T_1$ of the
  reprocessing surface in envelopes lacking outflow cavities. $T_1$ is
  computed as a function of outer disk period period using the opacity
  model of \cite{2003A&A...410..611S}.  Assumed parameters are
  $M=0.5\Msun$, $\dot{M} = 3\times 10^{-6}\Msun$ yr$^{-1}$,
  $T_\star=4500$ K, and solar metallicity; however, equation
  (\ref{Reprocessing}) permits this diagram to be rescaled for other
  parameters.  (The density dependence of evaporation temperatures has
  been suppressed for simplicity; lines here are drawn for $10^{-10}$ g
  cm$^{-3}$.)  Upper panel: effective opacity $\bar{\kappa}$ of cool
  dust to optically thin thermal radiation from dust of temperature
  $T_1$.
\label{reproc} }
\end{figure*}

Given a optically thin emission at the reprocessing temperature $T_1$,
the effective opacity for the second reprocessing event is given by 
\begin{equation}\label{Second-reproc}
\bar{\kappa} = \frac{\int  \kappa_{\nu2}\kappa_{\nu1} B_\nu(T_1)}
{\int  \kappa_{\nu1} B_\nu(T_1)}
\end{equation}
where $\kappa_{\nu1}$ is the opacity law at the reprocessing radius
and $\kappa_{\nu2}$ is that for cool dust near $R_d$.  Figure
\ref{reproc} shows $T_1$ and $\bar{\kappa}$ for a given model;
notably, $\bar{\kappa}\simeq 15$ cm$^{2}$ g$^{-1}$ for a wide range of
parameters, dropping only for especially large disk radii (and especially
low values of $\Sigsph(R_d)$). 

Integrating the flux through the disk at $R_d$ from the sky intensity
of reradiated light (assuming optically thin second reprocessing),
we find that the flux of this radiation at the disk radius $R_d$ is
given by equation (\ref{irradThick}) if
\begin{equation} \label{fFnoCavity}
\fF = 0.28 \bar{\kappa}\Sigsph  
\end{equation}
giving $\fF\simeq 1/16$ for typical parameters. Using this result to
calculate the irradiation temperature of an optically {\em thick} disk
in eq. (\ref{irradThick}), we find that it falls below the critical
temperature for local fragmentation at the outer disk edge only if

\begin{eqnarray}\label{PminIrradNoCavity}
{\rm period} &>& 
\frac{{7400f_p^{1.69}}}{f_j^{3/4}}
\left(\frac{\bar{\kappa}}{10\,\rm  cm^2\,g^{-1}}
\frac{\Sigmacl}{165\,\Msun\,\pc^{-2}}\right)^{3/4}
\nonumber \\ &~&\times 
\left(\frac{T_\star}{4500\,\rm K}\right)
\left(\frac{10\,\rm K}{T_c}\right)^{9/4}\yr
\end{eqnarray}
there; this is comparable to the characteristic maximum period in \S
\ref{Collapse}. 

Note we have taken $\varepsilon = 100\%$ here, since the absence of
the outflow cavity implies that all of the material is accreted onto
the star.  This raises the critical disk temperature to about 22 K.

\subsubsection{Outflow cavity} \label{Outflow}

A more realistic calculation must account for the region along the
axis that is cleared by the jet.  An outflow cavity warms the disk by
providing more direct illumination than is available otherwise.  To
calculate this we must specify the shape of the outflow cavity.

Note, first, that the wind ram pressure is expected generically to
vary as $1/(r^2 \sin^2 \theta)$ with distance and angle $\theta$ from
the axis \citep{1999ApJ...526L.109M}.  Inflow ram pressure scales as
$r^{-5/2}$ from the centrifugal radius to the edge of the inflow.  We
see on comparison that inflow dominates close to the star and near the
equator, and the wind dominates far from the star and near the axis.
By this reasoning, \cite{2000ApJ...545..364M} divided the initial core
into accreting and ejected angles depending on the velocity imparted
by the wind impulse to gas at the edge of the core.  Our parameter
$\varepsilon$ is simply the accreted mass as a fraction of the initial
mass.

By the same logic, gas that is not cast away by the wind is
destined to fall inward, and its motion is less and less affected by
the wind ram pressure as it does.  Therefore we are justified in
approximating the shape of the outflow cavity as an unperturbed
streamline in the infall solution of \cite{1984ApJ...286..529T}.  The
streamline in question is roughly the one that divides infall and
outflow in \cite{2000ApJ...545..364M}'s theory.  For the case of a
spherical initial core, then, the initial angle of this streamline is
given by 
\begin{equation}\label{OutflowCavity}
\cos \theta_0 = \varepsilon.
\end{equation}
(We shall assume a spherical core for the remainder of this section.) 
This streamline strikes the disk at a radius $\sin^2(\theta_0) R_d =
(1-\varepsilon^2) R_d$. 

\begin{figure*}
\centerline{\epsfig{figure=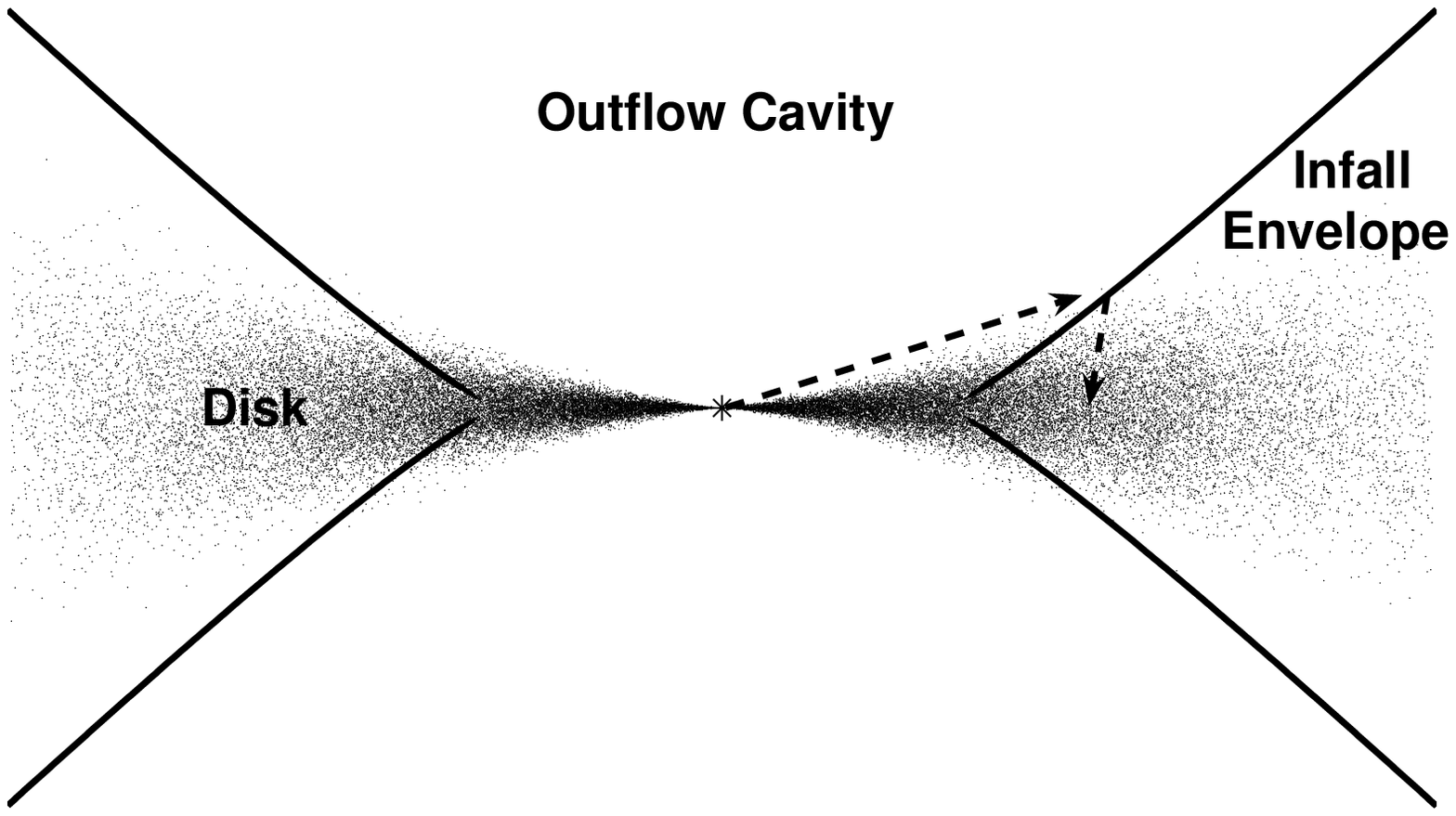,width=3.5in}} 
\caption[] {\footnotesize A schematic diagram for irradiation
  in the presence of an outflow cavity, in the theory of \S
  \ref{Outflow}.  The inflow envelope is excavated within a
  particular streamline (eq.~[\ref{OutflowCavity}], solid line);
  visible starlight is absorbed near this innermost streamline
  and  reprocessed into infrared light that illuminates the disk
  (dashed arrows). 
\label{reprocOutflow} }
\end{figure*}

Geometrically, a fraction $\varepsilon$ of the starlight strikes 
the cavity inner edge because the remaining $1-\varepsilon$ is cleared by
the outflow.  However, smaller $\varepsilon$ leads to a broader
outflow cavity, which causes a greater portion of the reprocessed
starlight to reach the disk.  
After performing ray-tracing calculations of reradiation from the
inner edge of the infall, using a geometry like that shown in figure \ref{reprocOutflow}, 
we find that $\fF$ is adequately
described by $0.1\varepsilon^{-0.35}$ in the relevant range
$20\%<\varepsilon<90\%$.   The infall envelope is translucent to this
reprocessed radiation, but our estimates indicate that shadowing of
the disk by the infall is insufficient to change this result
significantly. 

The disk's edge is too hot to fragment unless
\begin{equation}\label{PminIrradWithCavity}
{\rm period} > 2.4\times10^4  f_p^{9/8}
\left(\frac{60\%}{\varepsilon}\right)^{1.8}
\frac{T_\star}{4500\,\rm K}
\left(\frac{10\,\rm K}{T_c}\right)^{9/4}\yr. 
\end{equation}

Consider now the evolution of the effect of irradiation as the
accretion rate declines.  The critical disk midplane temperature,
eq. (\ref{T>Teff}), varies as $\dot{M}_{\rm acc}^{2/3}$.  The
radiative flux normal to the outer disk varies as the central
luminosity ($\propto \dot{M}_{\rm acc}$ for low-mass stars); if the
infall is optically thin then there is another factor of $\dot{M}_{\rm
acc}$ in the reprocessed fraction.  The flux therefore varies as
$\dot{M}_{\rm acc}^{1-2}$, implying a midplane temperature that varies
as $\dot{M}_{\rm acc}^{0.25-0.5}$.  The weaker dependence implies that
a disk which is stabilized by irradiation will remain so as the mass
supply wanes.   This might not hold if accretion were to stop suddenly
compared to the disk's viscous time, but this is not realistic. 

To summarize, {\em local fragmentation due to Toomre's instability is
quenched by irradiation in protostellar disks},
in all but exceptionally long-period systems whose orbital separations
are of order 1000 AU.  

However, disks fragment at smaller periods if they form from cores
that are warmer than 10 K or from more turbulent cores than were
described in \S \ref{Inits}.  Disks around massive stars fall in this
category \citep{2003ApJ...585..850M}, although their central
illumination is different than we have assumed.

\section{Global Instability}\label{global}

In light of the previous section, one expects irradiation to prevent
local fragmentation and possibly also to prevent the enhanced angular momentum
transport from self-gravity.  Note, however, that
\cite{2001ApJ...553..174G}'s results were for razor-thin disks, and
protostellar disks are rather thick in comparison.  Therefore, the
disk mass can become large enough to induce global instability.  As we
shall see, the threshold for this depends on the ability of local angular
momentum transport (e.g., due to MRI) to remove material from the disk. 

\cite{1989ApJ...347..959A} and \cite{1990ApJ...358..495S} describe a
global ``SLING'' instability in which the disk becomes massive enough
that the star is perturbed from the center of mass by the action of a
lopsided spiral.  \cite{1990ApJ...358..495S} provide the following
formula (their eq. [111]) for the criterion of instability:
\begin{equation}\label{SLING}
\frac{M_d}{M_d+M} = \frac{3}{4\pi} {\cal M}(Q)
\end{equation}
where 
\begin{eqnarray}\label{SlingCalM}
{\cal M}(Q) &\equiv& \int\left[x^4+4Q^2(x-x^{5/2})\right]^{1/2}dx \nonumber\\
&\simeq& 1 + \frac{8}{9} (Q-1)
\end{eqnarray}
where the approximation is exact in the limits $Q\rightarrow 1$ and
$Q\rightarrow \infty$ and holds within $2.3\%$ for all $Q>1$.  When it
sets in, this instability is thought to transport material rapidly
enough to prevent fragmentation \citep[][see however
\citealt{1994MNRAS.269..837B}]{1998ApJ...504..945L}.

We shall not posit $Q=1$, because we have seen in \S \ref{irrad} that
irradiation tends to warm the disk so that $Q>1$. Instead, let us
assume $\Sigma\propto r^{-3/2}$ so that $M_d = 4\pi R_d^2 \Sigma$.  If
we approximate $\Omega^2\simeq G(M+M_d)/R_d^3$ and retain $H =
c_s/\Omega$, then (\ref{Q}) becomes
\begin{equation}\label{Qthick}
Q = 4 \frac{H}{r} \left(1+\frac{M}{M_d}\right). 
\end{equation} 
With equations (\ref{SLING}) and (\ref{SlingCalM}), this gives a
maximum stable disk mass as a function of $H/r$.  We find that $Q>1$
and $M_d<M$ requires $0.06<H/r<0.28$; within this range, we
approximate the exact solution of equations (\ref{SLING}),
(\ref{SlingCalM}), and (\ref{Qthick}): 
\begin{equation} 
\frac{M_d}{M}< 3.09\frac{H}{r}+0.13 
\end{equation} 
for global stability.  The approximation is valid within $1\%$. 

This can be compared to the mass accumulated,  which for $\Sigma\propto
r^{-3/2}$ is 
\begin{equation}
M_d = \frac{\dot{M}_{\rm acc}}{\Omega}
\frac{4}{3\alpha} \left(\frac{r}{H}\right)^2.
\end{equation} 
The comparison sets a minimum for $\alpha/\alpha_0$, where 
\begin{equation} \label{defChiD}
\alpha_0 \equiv \frac{\dot{M}_{\rm acc}}{M \Omega}. 
\end{equation} 
The theory of \S \ref{Inits} indicates that  
\begin{equation} 
\alpha_0 = 10^{-2.6} \frac{f_j^3}{f_p^{3/4}}
\left(\frac{60\%}{\varepsilon}\right)^2,   
\end{equation} 
independently of the physical scale of the parent cloud. 

Taken together, we find that the minimum $\alpha$ required to keep the
disk stable can be approximated (within 4.2\%)
\begin{equation} \label{alphaMinGlobal}
\alpha > 0.52 \left(\frac{r}{H}\right)^{2.76} \alpha_0. 
\end{equation} 
The instability is clearly very sensitive to the midplane temperature. 
Computing $H/r = c_s/v_k$ using the irradiation temperature model of \S
\ref{Outflow} (which gives $H/r\sim 0.11$ for fiducial parameters), we
find that stability requires 
\begin{eqnarray} \label{alphaMinFromCore}
\alpha &>&0.19 \frac{f_j^{1.62}}{f_p^{1.27}}
\left(\frac{60\%}{\varepsilon}\right)^{0.27}
\left(\frac{T_c}{10\,\rm K}\right)^{1.27}\nonumber\\ &~&\times 
\left(\frac{165\,\Msun\,\pc^{-2}}{\Sigmacl}\right)^{0.46}
\left(\frac{4500}{T_\star}\right)^{0.46}. 
\end{eqnarray}

It has been argued that local gravitational transport becomes
significant when $Q$ falls below $\sim 1.4$
\citep{1994ApJ...436..335L}.  In its absence, the disk is left with
magnetoturbulent (MRI) transport, which is thought to saturate at much lower
values of $\alpha$ in a weakly ionized medium. 

With eq.~(\ref{alphaMinFromCore}), this  indicates that global spiral modes
{\em can be} required for protostellar disks to process the mass
supply, depending on the infall parameters and on the detailed
irradiation temperature (which we have calculated only approximately).
Disks that are especially thick or whose parent cores are especially
slowly rotating ($f_j<<1$) will be stable at smaller values of
$\alpha$.

\section{Summary and Conclusions}\label{summaryconclusions}
\subsection{Summary}\label{summary}
We derived in \S \ref{Inits} a simple set of relations for the
smallest and most thermally supported collapsible objects, i.e.,
cores, that can exist at a given mean gas temperature in a parent
cloud of a given column density.  The cloud column sets both the scale
for the cores' bounding pressure, and the coefficient in the line
width-size relation for cloud turbulence.  This turbulence aids in
core support and increases the minimum core mass, while also
accelerating the accretion rate when cores do collapse (\S
\ref{Collapse}).  In addition, turbulence bestows cores with
rotational angular momentum that determines the characteristic scales
for protostellar accretion disks (\S \ref{CoreRot}).  This theory does
not account for the role of a static magnetic field, nor does it
prescribe the distribution of overpressures expected in a realistic
model for molecular cloud turbulence; nevertheless, it matches 
the observations of prestellar cores of a couple solar masses (\S
\ref{ObsComp}) with a minimum of free parameters.  In our model,
substellar cores would derive from regions of especially low
temperature and high degrees of overpressure.  High mass cores could
come from the opposite conditions or, more likely, derive the bulk of
their initial support from nonthermal pressure.  

Turning next to the fragmentation boundary as predicted even in
axisymmetric models of protostellar collapse, we found in \S
\ref{Impulsive} that the cores are typically subject to a phase of
prompt fragmentation according to \cite{2003ApJ...595..913M}'s
criterion.  This is a possible mechanism for the production of
relatively tight binary stars.  However, we note that this occurs
before most of the core mass has accreted; therefore these fragments
may be driven to merge by radiating angular momentum to the accretion
disk.
We also stress that \citeauthor{2003ApJ...595..913M}'s simulations
used an approximation to the thermal evolution of the collapsing core,
and that we and others have found the gas temperature to be a
controlling physical factor in fragmentation.  We therefore
do not consider it certain that binaries form by this mechanism. 

Examining the local gravitational stability of protostellar disks in
\S \ref{localcriteria}, we made extensive use of the simulations by
\cite{2001ApJ...553..174G} to identify a critical disk temperature
that divides accreting from fragmenting disks.  For disks heated
solely and locally by their own viscosity, we found in \S
\ref{self-luminous} that local fragmentation is impossible for disk
periods shorter than about 460 years (depending on the opacity law),
and that fragmentation is suppressed until much longer periods if the
disk is optically thin to its own cooling radiation.  This conclusion
is entirely consistent with the recent argument by
\cite{2004astro.ph..6469R} that the formation of giant planets by
direct disk instability \citep{1997Sci...276.1836B} is impossible on
thermodynamic grounds.  Moreover, when we account for heating by the
star's accretion luminosity (reprocessed by the infall envelope), we
find that the local instability is entirely quenched out to very long
periods, $\sim 10^{4.1}$ years.  A decline in
the accretion rate does not alter this conclusion.

Lastly, we consider the onset of global instability due to the
accumulation of a finite disk mass.  Applying the instability
criterion of \cite{1990ApJ...358..495S}, we find that the intrinsic
viscosity parameter $\alpha$ (e.g., due to MRI) must exceed 
roughly 0.2
in order to prevent global instability.  
This makes global instability a primary candidate for angular momentum
transport within protostellar disks.
When it does occur, this instability is likely to saturate and provide
rapid angular momentum transport rather than fragmentation
\citep{1998ApJ...504..945L}.

\subsection{Implications for Numerical Simulations of Multiple Star
  Formation}\label{NumericalImplications}

 Thermal evolution plays a critical role in determining the onset and
outcome of gravitational stability in disks
\citep{1964ApJ...139.1217T, 1981Icar...48..377C,
1991ApJ...382..530T,1994ApJ...422..850T,
1998ApJ...504..468P,1999numa.conf..177P, 2000ApJ...529.1034P,
2001ApJ...553..174G,2003ApJ...590.1060P}.  We find that local disk
fragmetation is quenched by stellar irradiation except at very large
periods (\S~\ref{irrad}; see also \citealt{1997ApJ...474..397D}) and,
even in the absence of irradiation, inhibited for short periods
($<460$ years) by viscous heating (\S \ref{self-luminous}).

These observations are quite important for the interpretation of
large-scale numerical simulations designed to predict star
formation.  Currently, only a few multidimensional calculations have
accounted for radiation transfer during collapse
\citep{2000ApJ...528..325B}, and then only in an approximate manner.
No dynamical simulations have yet included disk irradiation due to starlight
that is reprocessed by the infall envelope.  Disk fragmentation in
these simulations is unlikely to be realistic. 

For instance, three quarters of the brown dwarfs formed in the
simulations of \cite{2002MNRAS.332L..65B} arise via disk
fragmentation.  Since these simulations employ a barotropic equation of
state, we predict that these brown dwarfs would not form if stellar
irradiation were accounted for. 

We have demonstrated in \S \ref{Impulsive} that initial conditions for
most collapsing cores are favorable for the prompt fragmentation
simulated by \cite{2003ApJ...595..913M}.  However, since these were
also performed with a barotropic equation of state, and since they do
not extend long enough to determine whether the fragments will merge,
we consider them provisional and await radiation transfer calculations
of the same parameter space.

\subsection{Conclusions: Stellar Binaries and the Brown Dwarf
  Desert}\label{conclusions} 

Our foremost conclusion is that stellar and substellar companions are
not produced by disk fragmentation,   
except possibly in an early phase of core collapse (although we
argue this is unlikely: see \S \ref{Impulsive}) or at very large
periods of $\sim 10^{4.1}$ years (\S \ref{irrad}). 
This indicates that turbulent perturbations beyond rotation and
compression are responsible for the production of binary stars
\citep{2003RMxAC..15...92K,2002MNRAS.332L..65B}.  
Since companions formed in disk instabilities would be initially of
substellar mass, this closes a formation channel for substellar
companions
over a wide range of orbital periods. 
Observations \citep[e.g.,][]{2004AJ....127.2871M} show a
marked dearth of companions in the mass range between stars and
planets -- a brown dwarf desert.  \cite{2002MNRAS.330L..11A} have
ascribed this to inward disk migration that destroys brown dwarfs.
However this presupposes that they can form within disks; here we have
argued that they cannot.

\cite{2004ApJ...604..284B} argue from their observations of disk
accretion onto free-floating brown dwarfs that these must form in
isolation rather than from fragmentation in a 
disk or in a collapsing turbulent core.  Our theoretical results
also argue against a disk origin for these objects.

\acknowledgments
We thank Jonathan Williams, Ue-Li Pen, and Fred Adams for helpful and
encouraging discussions; Mary Barsony and Sukanya Chakrabarti for
useful comments; Roman Rafikov for pointing us to his preprint; and
Dmitry Semenov for explanations concerning the
\cite{2003A&A...410..611S} opacity model.  
Comments from our anonymous referee have been very helpful in
improving the clarity of the paper. 
CDM's research is funded by NSERC and the Canada Research Chairs
program.  YL is supported by an NSERC senior CITA postdoc.

\appendix
\section{Specific angular momentum distribution of prestellar cores}
\label{Appendix:j}

We now compute the angular momentum distribution quoted in equations
(\ref{j}) and (\ref{fj}) according to the model for core rotation
described in \S \ref{CoreRot}.  
Specifically, we assume in the present calculation a spatially
homogeneous velocity fields $\mathbf v$ whose components are
uncorrelated Gaussian fields obeying a line width-size scaling like
that in equation (\ref{Larson}).  The effect of gas pressure, which
introduces correlations by inhibiting compression, is discussed
separately in \S \ref{CoreRot}.  As a consequence of homogeneity,
we assume no correlation between $\mathbf v$ and the core density field
$\rho(r)$. See \S \ref{PrevThy} for a discussion of these and
alternative assumptions. 

Since we are interested in the specific angular momentum normalized to
$R \sigmaNT(R)$, we compute $j$ and $\sigmaNT(R)$ separately.

\subsection{Angular momentum}\label{Appendix:L}

To satisfy the line width-size relation in this model, we must impose 
that the velocity difference between the two points scales as a square
root of the distance between the points.  More precisely,
\begin{equation}
\langle [v_i(r_1)-v_j(r_2)]^2\rangle=k|{\mathbf r}_1-{\mathbf r}_2| \delta_{ij}
\label{difference}
\end{equation}
where we assume that $|r_1-r_2|\ll v^2/k$ if 
$v=\langle |{\mathbf v}|^2 \rangle^{1/2}$ is the characteristic
turbulent velocity.  
Here, angle brackets denote an ensemble average over
realizations of the turbulent velocity field, or equivalently, over
the location of the core within this field. 
Equation (\ref{difference}) is assumed to hold for each velocity
component $v_i$ independently.
Expanding equation (\ref{difference}), 
\begin{equation}
\langle v_i(r_1)v_j(r_2)\rangle=\left({1\over 3}v^2-{1\over 2}
   k|{\mathbf r}_1-{\mathbf r}_2| 
\right)\delta_{ij}.
\label{correlation}
\end{equation}

Skipping to the result, the root mean square angular momentum
of a core in this velocity field is given by
\begin{equation}
\langle L^2\rangle^{1/2}=2\pi k^{1/2}R^{9/2}\rho_0 \left[F(\bar{\rho})
\right]^{1/2},
\label{L}
\end{equation}
where $R$ is the core radius and $\rho_0$ is 
a reference density; 
$\bar{\rho}(\bar{r})=\rho(r)/\rho_0$ is the spherically
symmetric normalized density profile expressed in terms of the
dimensionless radius $\bar{r}=r/R$, and the form factor $F$ is given
by the following double integral:
\begin{eqnarray}
F&=&\int_0^1\int_0^1 d\bar{r}_1 d\bar{r}_2 \bar{\rho}(\bar{r}_1)
\bar{\rho}(\bar{r}_2) \bar{r}_1\bar{r}_2 ({\bar{r}_1^2+\bar{r}_2^2})^{5/2}
\times\nonumber\\
& &~~~~~~\left\{{1\over 5}\left[(1+q)^{5/2}-(1-q)^{5/2}\right] 
\right. \nonumber \\ 
& &~~~~~~~~~~ \left.-  {1\over 3}\left[(1+q)^{3/2} -(1-q)^{3/2}\right]\right\}.
\label{F}
\end{eqnarray}
Here $q=2r_1r_2/(r_1^2+r_2^2)$.

Evaluated over a Bonnor-Ebert sphere, 
$F^{1/2} = 0.0294$. Such spheres obey $M=0.735 \rho_0 R^3$, so 
that 
\begin{equation} \label{jfromF}
\langle j^2 \rangle^{1/2} = \frac{\langle {L^2}\rangle^{1/2}}{M} = 0.251
k^{1/2} R^{3/2}.
\end{equation}

The derivation is as follows.  The total angular momentum is given by
\begin{equation}
L^2=L_x^2+L_y^2+L_z^2.
\end{equation}
By spherical symmetry the components of angular momentum
are uncorrelated; this can be checked directly.  Therefore,
\begin{equation}
\langle L^2\rangle=3\langle L_z^2 \rangle=3\langle (L_1-L_2)^2\rangle,
\end{equation}
where
\begin{eqnarray}
L_1&=&\int d^3{\mathbf r}\rho(r)xv_y,\\
L_2&=&\int d^3{\mathbf r}\rho(r) yv_x\label{l1l2}
\end{eqnarray}
But $\langle L_1 L_2\rangle=0$ and, by spherical symmetry,
$\langle L_1^2\rangle=\langle L_2^2\rangle$. Therefore,
\begin{eqnarray}
\langle L^2\rangle&=&6\langle L_1^2 \rangle\\
&=&6\int\int d^3 {\mathbf r}_1 d^3{\mathbf r}_2\ \rho(r_1)\rho(r_2)x_1 x_2 \langle
v_y({\mathbf r}_1)v_y({\mathbf r}_2)\rangle.\nonumber
\end{eqnarray}

We can now use Eq.~(\ref{correlation}) and evaluate the integral above.
First, note that only the second term on the right-hand side of 
Eq.~(\ref{correlation}) contributes to the integral. Second, because
of the spherical symmetry we can substitute $x_1x_2$ by $({\mathbf r}_1
{\mathbf r}_2)/3$ in the integrand.
Thus we have for the total angular momentum:
\begin{equation}
\langle L^2 \rangle=
-k\int\int d^3{\mathbf r}_1 d^3{\mathbf r}_2 \rho(r_1) \rho(r_2) ({\mathbf r}_1{\mathbf r}_2)
          |{\mathbf r}_1-{\mathbf r}_2|.
\label{totalL}
\end{equation}
The integrand in Eq.~(\ref{totalL}) depends only on the magnitudes of
${\mathbf r}_1$ and ${\mathbf r}_2$, and on the angle $\theta$ between them.
 Therefore, the 
other three free variables are integrated out, and we are left with the 3-d
integral:
\begin{eqnarray}
\langle L^2 \rangle&=&-k\int dr_1 4\pi r_1^2\rho(r_1)\times\nonumber\\
& &\int\int dr_2 2\pi d\cos\theta
 \cos\theta \left[\rho(r_2) r_1 r_2
  \sqrt{r_1^2+r_2^2}\nonumber\right. \\&&~~~~~~~~
  ~~~~~~~~~~~~~~~~~~~~~~~~~~
\left. \sqrt{1-q\cos\theta} \right],
\label{Ltotal}
\end{eqnarray}
where
\begin{equation}
q={2r_1r_2\over r_1^2+r_2^2}.
\end{equation}
The angular integration in Eq.~(\ref{Ltotal}) can be performed analytically,
most simply by using $\sqrt{1-q\cos\theta}$ as an integration variable.
The result of this calculation is presented in
Equations (\ref{L}) and (\ref{F}).

\subsection{One-dimensional velocity dispersion}\label{Appendix:sigma}
We now compute the one-dimensional velocity dispersion:
\begin{eqnarray}\label{sigmadef} 
\sigma^2 &=& \int d^3{\mathbf r}\ \tilde{\rho} \left[v_x({\mathbf r}) - \bar{v}_x\right]^2  
\nonumber\\ &=& \int d^3{\mathbf r}\ \tilde{\rho}  \left[v_x({\mathbf r})^2  -
2 v_x({\mathbf r}) \bar{v}_x + \bar{v}_x^2 \right]
\nonumber \\ &=& \left[\int d^3{\mathbf r}\ \tilde{\rho} v_x({\mathbf r})^2 \right] -
  \bar{v}_x^2
\end{eqnarray}
where $\tilde{\rho}\equiv \rho/M$ and $\bar{v}$ denotes the center of
mass velocity:
\begin{equation}
\bar{v} = \int d^3{\mathbf r}\ \tilde{\rho} v({\mathbf r})
\end{equation}

In equation (\ref{sigmadef}), the first term reduces after ensemble
averaging to $v^2/3$, according to equation (\ref{correlation}).
Likewise for the second term, 
\begin{eqnarray}\label{vcmsquared}
\langle \bar{v}_x^2 \rangle &=& \int\int d^3{\mathbf r_1}d^3{\mathbf r_2}\
\tilde{\rho}({\mathbf r_1})\tilde{\rho}({\mathbf r_2}) \langle
v_x({\mathbf r_1})v_x({\mathbf r_2})\rangle 
 \\ \nonumber
&=& \int\int d^3{\mathbf r_1}d^3{\mathbf r_2}\ 
\tilde{\rho}({\mathbf r_1})\tilde{\rho}({\mathbf r_2}) \left(\frac{1}{3}v^2 -
  \frac{1}{2}k |{\mathbf r_1}-{\mathbf r_2}|\right). 
\end{eqnarray}
The constant terms cancel in the ensemble average of
eq. (\ref{sigmadef}), leaving 
\begin{eqnarray} \label{<sigma^2>}
\langle \sigma^2 \rangle &=&    \frac{1}{2}k 
\int\int d^3{\mathbf r_1}d^3{\mathbf r_2}\ 
\tilde{\rho}({\mathbf r_1})\tilde{\rho}({\mathbf r_2}) |{\mathbf r_1}-{\mathbf r_2}|
\nonumber \\ &=& \frac{1}{2}k \int 4\pi r_1^2 dr_1 \tilde{\rho}(r_1) 
\int 2\pi r_2^2 \tilde{\rho}(r_2) \sqrt{r_1^2+r_2^2} \times 
\nonumber \\ && ~~~~~~~ ~~~~
 \int d\cos\theta \sqrt{1-q\cos\theta}
\nonumber \\ &=& 
\frac{8\pi^2}{3} k R \int\int d\bar{r}_1 d\bar{r}_2 \ \bar{r}_1^2
\bar{r}_2^2 \tilde{\rho}(\bar{r}_1) 
 \tilde{\rho}(\bar{r}_2) \sqrt{\bar{r}_1^2+\bar{r}_2^2} \times 
\nonumber \\ &&~~~~~~~~~~~\left[(1+q)^{3/2} - (1-q)^{3/2}\right]. 
\end{eqnarray}

Evaluated over a Bonnor-Ebert sphere,
\begin{equation} \label{<sigma>_rms}
\langle \sigma^2\rangle^{1/2} = 0.607 k^{1/2} R^{1/2}, 
\end{equation}
and in combination with equation (\ref{jfromF}), 
\begin{equation}\label{<j>from<sigma>R}
\langle j^2 \rangle^{1/2} = 0.414 \langle \sigma^2 \rangle^{1/2} R. 
\end{equation}

Since the velocity distribution is Gaussian, all of its linear moments
(e.g., $L_x$ and $\sigma$) are Gaussian distributed.  The cumulative 
probability distribution of $j$, ${\cal P}(<j)$, is therefore 
Maxwellian with a width fixed by equation (\ref{<j>from<sigma>R}):
\begin{equation}\label{jdistribution}
\frac{d {\cal P}(<j)}{d (j/\langle j^2 \rangle^{1/2})} = 
  3\sqrt{\frac{6}{\pi }} \frac{j^2}{\langle j^2
  \rangle} \exp\left(-\frac{3 j^2}{2\langle j^2 \rangle}\right)
\end{equation}  
A log-normal fit to this distribution is $\log_{10}{j/\langle
j^2\rangle^{1/2}} = {-0.080\pm0.52}$, but
this is not especially accurate.  A more faithful representation is 
$\log_{10}{j/\langle j^2\rangle^{1/2}} = -0.088^{+0.16}_{-0.49}$.

\end{document}